\documentclass[12pt]{article}
\usepackage{epsfig}
\newcommand{\be}{\begin{equation}}
\newcommand{\ee}{\end{equation}}
\newcommand{\bea}{\begin{eqnarray}}
\newcommand{\eea}{\end{eqnarray}}
\newcommand{\nref}[1]{(\ref{#1})}
\newcommand{\la}[1]{ \label{#1}}
\newcommand{\lam}{\lambda}
\newcommand{\eps}{\epsilon}

\begin{document}

\begin{titlepage}
\begin{center}
\hfill   hep-th/0502086

\vskip 1cm

{\LARGE {\bf Deforming field theories with $U(1)\times U(1)$ \\ ~ \\
global
  symmetry and their gravity duals
  }}

\vskip 1.5cm

{\large  Oleg Lunin and Juan Maldacena }

\vskip 2cm

   School of Natural Sciences, Institute for Advanced Study,
Princeton, NJ 08540

\vskip 3.5cm

\vspace{5mm}

\noindent

{\bf Abstract}

\end{center}

 We find the gravity dual of a marginal
deformation of ${\cal N}=4$ super Yang Mills, and discuss some of
its properties.  This deformation is intimately connected with an
$SL(2,R)$ symmetry of the gravity theory. The $SL(2,R)$
transformation enables us to find the solutions in a simple way.
These field theory deformations, sometimes called $\beta$ deformations,
 can be viewed as arising from a
star product. Our method works for any theory that has a gravity
dual with a $U(1)\times U(1)$ global symmetry which is realized
geometrically. These include the field theories that live on D3
branes at the conifold or other toric singularities, as well as
their cascading versions.

\vskip 4.5cm

{\tt lunin@ias.edu, malda@ias.edu}

\end{titlepage}

\newpage

\section{Introduction}
\renewcommand{\theequation}{1.\arabic{equation}}
\setcounter{equation}{0}

The gauge theory/gravity duality (or AdS/CFT)
relates field theories to gravitational theories
with particular boundary conditions \cite{jm,gkp,wittenhol}.
Modifications in the boundary conditions
correspond to changes in the field theory lagrangian.
In this paper we consider a specific type of deformation of the
field theory lagrangian  and its corresponding gravity dual. These are sometimes
called $\beta$ deformations.
One example of the class of deformations we study is the
marginal  \cite{marginal}\cite{rlms} deformation of
${\cal N}=4 $ Yang-Mills theory to ${\cal N}=1$ which preserves a $U(1) \times U(1)$
non-R-symmetry\footnote{
By a non-R-symmetry we mean a symmetry that leaves the ${\cal N} =1$
supercharges invariant. In addition, ${\cal N}=1$ superconformal theories
 have a $U(1)_R$ symmetry.}. In general, we consider $U(N)$ field theories with
 a $U(1)\times U(1)$ global symmetry.  The deformation we study can be
viewed as arising from a new definition of the product of fields
in the lagrangian \be \label{deform} f * g \equiv e^{ i  \pi
\gamma (Q^1_f Q^2_g - Q^2_f Q^1_g) } f g \ee where $fg$ is an
ordinary product and $(Q^1,Q^2)$ are the $U(1) \times U(1)$
charges of the fields. Though this prescription is similar in
spirit to the one used to define non-commutative field theories
\cite{cds,nsew}, the resulting theory is an ordinary field theory.
All that happens is that (\ref{deform}) introduces  some phases in
the lagrangian. For example, in the ${\cal N}=4\to {\cal N} =1 $
deformation we mentioned above this results in the change of
superpotential \be  \la{supotch} Tr( \Phi_1 \Phi_2 \Phi_3 -\Phi_1
\Phi_3 \Phi_2 ) \to Tr( e^{ i \pi \gamma } \Phi_1 \Phi_2 \Phi_3 -
e^{-i \pi \gamma } \Phi_1 \Phi_3 \Phi_2 ) \ee
We shall call this the ``$\beta$-deformed'' theory.
Suppose that we know
the gravity dual of the original theory and that this geometry has
two isometries associated to the two $U(1)$ global symmetries.
Thus the geometry contains a two torus. The gravity description of
the deformation (\ref{deform}) is surprisingly simple. We just
need make the following replacement \be \label{gravdeform} \tau
\equiv B + i \sqrt{g} \longrightarrow \tau_\gamma = {\tau \over 1 + \gamma
\tau } \ee in the original solution, where $\sqrt{g}$ is the
volume of the two torus.
 We can view (\ref{gravdeform}) as a solution generating
transformation. Namely,   we reduce the ten dimensional theory to eight dimensions
on the two torus. The eight dimensional   gravity theory
is invariant under $SL(2,R)$ transformations
acting on $\tau$. The deformation (\ref{gravdeform}) is one particular element
of $SL(2,R)$. This particular element has the interesting property that it
produces a non-singular metric if the original metric was non-singular.
The $SL(2,R)$ transformation could only produce
 singularities when $\tau \to 0$. But  we see from (\ref{gravdeform}) that
$\tau_\gamma = \tau + o(\tau^2) $ for small $\tau$. Therefore,
near the possible singularities the ten dimensional metric is
actually same as the original metric, which was non-singular by
assumption.

In the rest of this paper we explain this idea in more detail.
Section two is devoted to a detailed explanation of the action of
the $SL(2,R)$ transformation \nref{gravdeform} which plays a
central role in this paper. We explain how it is associated to
\nref{deform} and point out its intimate connection with
non-commutative theories. In section three we apply this method to
obtain the gravity solution for the marginal deformation of
${\cal N}=4$ super-Yang-Mills. We also discuss various features
that arise at rational values of the parameter $ \gamma$ in \nref{supotch}.
These field theories were studied in
\cite{berenstein1,berenstein2,berenstein3,ofer,doreysduality,doreynsfive,doreycoulomb,benini,hollowood,niarchos}.
In section four we discuss more general cases. We discuss a marginal
deformation of the conifold theory studied by Klebanov and Witten
\cite{ikew}. We also generalize this to marginal deformations of
more general theories
based on toric manifolds \cite{italian,francoetal,MartSparks}.
 We also point out that this method applies also to
non-conformal theories, so it can be applied to obtained a
deformed version of the cascading theory of Klebanov and Strassler
\cite{ikms}. I section five we present a generalization of this method to
theories with three $U(1)$ symmetries, which leads to marginal deformations
of $AdS_4 \times S^7$.

In appendix A we give more details on the action of the solution generating transformation
and we give the detailed metrics for various cases mentioned in the main text.
In appendix B we give present
classical string solutions that are associated to BPS states in the
$\beta$-deformation of ${\cal N}=4$ super Yang-Mills for rational deformation parameters.

\section{An SL(2,R) transformation}
\renewcommand{\theequation}{2.\arabic{equation}}
\setcounter{equation}{0}

Let us consider a string theory background with two $U(1)$
symmetries that are realized geometrically. Namely there are two
coordinates $\varphi_1,\varphi_2$ and the two $U(1)$ symmetries act on
these two coordinates as shifts of $\varphi_i$.
Then we will have a two torus parametrized by $\varphi_i$, which, in
general, will be
fibered over an eight dimensional manifold.
 A simple  example is the metric
of $R^4$
 \be \la{rfour}
 ds^2 = d\rho_1^2 + d\rho_2^2+ \rho_1^2 d\varphi_1^2
+ \rho_2^2 d\varphi_2^2 \ee
As this example shows, the two torus could contract to zero size at some points but
nevertheless the whole manifold is non-singular.

When we compactify a closed string theory on a two torus the resulting eight dimensional
theory has an exact $SL(2,Z) \times SL(2,Z)$ symmetry which acts on the complex structure
of the torus and on the parameter\footnote{ Throughout this paper we
 are setting $\alpha'=1$ and we are using
a normalization for the $B$ field such that its period is $B_{12} \sim B_{12}+1$. }
\be
\tau = B_{12} + i \sqrt{g} \la{taudef}
\ee
where $\sqrt{g}$ is the volume of the two torus in string metric.
The $SL(2,Z)$ that acts on the complex structure will not play an important role and
we forget about it for the moment.
At the level of supergravity we have an $SL(2,R) \times SL(2,R)$  symmetry. This is not
a symmetry of the full string theory. The  $SL(2,R)$ symmetries of supergravity
can be used as solution generating transformations. This is a well known  trick which
was used to generate a variety of solutions in the literature \cite{hor}.
The $SL(2,R)$ symmetry that plays a central role in this paper is the one acting as
\be \la{sltwo}
\tau \to \tau' = { \tau \over 1 + \gamma \tau}
\ee
where $\tau$ is given by \nref{taudef}. Of course, we can also think of \nref{sltwo} as
the result of doing a T-duality on one circle, a change of coordinates, followed by another
T-duality. When $\gamma$ is an integer \nref{sltwo}
is an $SL(2,Z)$ transformation, but for general $\gamma$ it is not. This transformation
generates a new solution. For example, applying this to \nref{rfour}
we get\footnote{This formula is OK for the bosonic string. For the superstring
we should replace $\gamma \to 2 \gamma$ if we want an integer gamma to correspond
to an $SL(2,Z)$ symmetry. This is due to the fermion periodicity conditions. }
\be \la{exsltwo}
\tau = i \rho_1 \rho_2  \to  \tau' = { \gamma \rho_1^2 \rho_2^2 \over1 + \gamma^2 \rho_1^2
\rho_2^2} + i { \rho_1 \rho_2 \over  1 + \gamma^2 \rho_1^2 \rho_2^2 }
\ee
The metric after the transformation \nref{exsltwo} is
\bea \label{rfoursol}
ds^2 &=&  d\rho_1^2 + d\rho_2^2+ { 1 \over  1 + \gamma^2 \rho_1^2 \rho_2^2}
( \rho_1^2 d\varphi_1^2
+ \rho_2^2 d\varphi_2^2)
\\
B_{12}& =& { \gamma \rho_1^2 \rho_2^2 \over1 + \gamma^2 \rho_1^2
\rho_2^2} \nonumber
\\
e^{2 \phi} &= & e^{2 \phi_0} {1 \over  1 + \gamma^2 \rho_1^2
\rho_2^2} \nonumber
\eea
where the change in the dilaton is due to the fact that the $SL(2,R)$ transformation leaves
the eight dimensional dilaton invariant, not the ten dimensional one. ($\phi_0$ is the
original ten dimensional dilaton).
This type of background is similar to the flux branes \cite{fluxbranes}\footnote{
In fact if we apply this trick to the initial metric $ds^2 = dr^2 + r^2 d\varphi_1^2
+ d\varphi_2^2$ we get a solution where the $H^{NS}_3$ field  has a nonzero value at the
origin. These solutions were studied in \cite{fluxbranes}.}.
It might be possible to quantize strings exactly in these backgrounds, as it
was done in \cite{russo} for  similar cases.

Suppose that we start with a non-singular ten dimensional geometry. After applying \nref{sltwo},
when
is the geometry  non-singular?. Let us assume that
in the original ten dimensional geometry the $B$ field goes to zero when $\tau_2 \to 0$.
This will be obviously true if the original $B$ field is zero, as in the example
we considered above, based on
\nref{rfour}.
Note that in order for the original solution to be non-singular we only need that
$\tau_1 $ goes to an integer when $\tau_2\to 0$. We might run into trouble if there are
different regions where $\tau_2\to0$ and $\tau_1 $ goes to different integers. In these
cases, we produce singularities and
our method cannot be applied\footnote{ An example were we get a singular solution arises
when we consider an NS five brane and we perform this procedure based on
$U(1)\times U(1) \in SO(4)$ acting on the transverse dimensions.}.
Under these assumptions the  transformation \nref{sltwo}
will give us a geometry that is non singular.
 The reason is that the only points where
we could possibly introduce a singularity by performing an $SL(2,R)$ transformation is
where the two torus shrinks to zero size. In this case $\tau_2 \to 0 $ and by assumption
we also have $\tau_1\to 0$.
If $\tau$ is small, then $\tau'$ becomes equal to $\tau$ and  the region near
the possible singularity becomes equal to what it was before the transformation. Thus
 the metric remains non-singular. For initial configurations with non-zero NS or RR
 fieldstrength
 the analysis of regularity is a bit longer and is discussed further in appendix A.
 Notice that \nref{sltwo} is the only
$SL(2,R)$ transformation with this property. In a more general situation when the
torus is fibered over the remaining eight dimensions a similar argument goes
through (see appendix A for more details on the general conditions).
This argument also shows that the topology of the solution remains the same.
The detailed formulas for the action of the $SL(2,R)$ transformation in the most
general case are given in appendix A.

Let us consider a D-brane on the original background
that is invariant under both $U(1)$ symmetries. Such a brane will be left invariant
under the action of \nref{sltwo}. In other words, there is a corresponding brane
on the new background.
We now ask the question:
 what is the theory on this brane in the new background?
We conjecture  the following answer.
Suppose that the original brane, on the original background,
gave rise to a certain open string field theory.
Then the open string field theory on the brane living
 on the new background is given by changing
the start product
\be \la{sft}
f *_\gamma  g   \equiv e^{   i\pi \gamma (Q^1_f Q^2_g - Q^2_f Q^1_g) } f *_0 g
\ee
where $*_0$ is the original star product and $Q^i_{f,g}$ are the $U(1)$ charges
of the fields $f$ and $g$.
The basic idea leading to this conjecture is the following. In
\cite{nsew} it was pointed out
that in the presence of a $B$ field the open string field theory
is defined in terms of an open string metric and non-commutativity parameter
\be \la{openmet}
G^{ij}_{open} + \Theta^{ij} = \left( {1\over g + B }\right)^{ij} \sim { 1 \over \tau}
\ee
where the last expression is schematic. Note that under
the transformation \nref{sltwo} $1/\tau \to 1/\tau' = 1/\tau + \gamma$. All that happens
in \nref{openmet} is that
  we introduce a non-commutativity parameter
$\Theta^{12} = \gamma$. The open string metric remains the same.
The reason we called this a ``conjecture" rather than a derivation is
that
\cite{nsew} considered a  constant metric and $B$ field while here we are
applying their formulas in a case where these fields vary in spacetime.

Let us now consider branes sitting at the origin. In general, the ``origin" is
 the point where both circles shrink to zero size\footnote{
If none of the circles shrinks, then we have the non-commutative theories considered
in \cite{nsew}. If one circle shrinks and the other does not shrink then we are
lead to the dipole theories studied in \cite{ganor}. These arise after taking the metric
 $ds^2 = d\rho^2 + \rho^2 d\varphi_1^2 + d \phi_2^2$, putting a D-brane at the origin which is
 extended in $\varphi_2$  and
performing the transformation \nref{sltwo}.   }.
Notice that for this brane the transformation \nref{sft} does {\it not} lead to a
non-commutative
field theory at low energies, since the $U(1)$ directions are not along its
worldvolume but they are global symmetries of the field theory.
The net effect of \nref{sft} for the field theory living on a brane  is to
introduce certain phases in the lagrangian according to the rule in \nref{deform}.
In other words,  starting with the low energy conventional field theory living
on the brane, we obtain another conventional field theory with some phases in
the lagrangian according to \nref{deform}.  Viewing the deformation as
a $*$ product allows us to show that all planar diagrams in the new theory are
the same as in the old theory \cite{filk}. Then, for example, if the original theory was
conformal, then this is a marginal deformation to leading order in $N$.
The reader might be bothered by the fact that we are using the arguments in
\cite{nsew} to derive the theory on a brane sitting at a point were $\sqrt{g} =0$.
One indirect argument for our procedure is the following. Start with a $D(p+2)$ brane
anti-brane system, wrapped on the two torus,
 with a magnetic  flux of their worldvolume $U(1)$s on the two torus
 so that we have  net  $Dp$ brane charge. These branes can annihilate via
 tachyon condensation \cite{sentachyon} to form the $Dp$ brane at the origin.
 The brane anti-brane system
 can be located far from the origin. The process of tachyon condensation is insensitive
 to the $\Theta$ parameter, so it proceeds in the same way in the theory after the
 $SL(2,R)$ transformation as in the theory before the transformation. The net result
 is that we obtain the same field theory on the $Dp$ brane at the origin, but
 with the extra phases \nref{deform}.

Notice that these arguments are completely general and can also be applied for
toric singularities, such as the conifold. In those cases we use the
$U(1)$ symmetries of the toric manifold and apply \nref{sltwo}. The field
theories on D-branes living at the singularity are deformed according to \nref{deform}.

The description in terms of this $SL(2,R)$ transformation \nref{sltwo} and its
associated modification of the theory \nref{sft} unifies the conceptual description
of non-commutative field theories \cite{cds,nsew} and dipole theories \cite{ganor} with
the so called $\beta$ deformations of field theories, which are the main subject of
this paper.

There are some special features that occur when $\gamma = m/n$, with $m,n$ coprime.
 In this case it is
possible to do a further $SL(2,Z)$ transformation on $\tau'$ to give
\bea
\tau'' &=& {a \tau' + b \over c\tau' + d} = { a \tau' + b \over -m \tau' + n}
= { 1\over n^2} \tau + { b \over n} \la{taudp}
\\
& & {\rm with} ~~~a n + m b =1 \la{diof}
\eea
Since $m,n$ are coprime, there is a solution to \nref{diof}.
The final expression for $\tau''$ \nref{taudp}
is precisely that of a
$Z_n \times Z_n$ orbifold  with discrete torsion of the original torus.
The discrete torsion is simply the last
term in \nref{taudp}.
Up to $SL(2,Z)$ transformations the final form of $\tau''$ is independent
of our choice of $b$.
The fact that D-branes in orbifolds with discrete torsion are related to deformations
of the field theory of the form \nref{deform} was pointed out in
\cite{douglasfiol} and further explored in \cite{berenstein1}.

An interesting question is whether the deformation that we are doing preserves
supersymmetry. In principle we can perform this transformation independently of whether
we break or preserve supersymmetry, but sometimes we are interested in the ones
that preserve it.
If the original ten dimensional background is supersymmetric under
a supersymmetry that is invariant under $U(1)\times U(1)$, then the deformed
background will also be invariant under this supersymmetry.
As an example, let us start with $R^4 \times R^6$.
In $R^6$ we choose complex coordinates $z^i$. Then we choose a subgroup
 $U(1)\times U(1) \subset SU(3)
\subset SO(6)$. Since these $U(1)$ symmetries act geometrically, we can apply
our construction. The resulting solution preserves ${\cal N}=2$ supersymmetry in
four dimensions. The interested
reader can find its explicit form in appendix A.
A D3-brane at the origin leads to
  an ${\cal N}=1$ theory. Actually, we obtain the marginal deformation
\nref{supotch}.   We discuss this
case more explicitly in the next section.
More generally we can imagine a toric singularity with three $U(1)$ symmetries.
We can pick two  $U(1)$ symmetries that -- leave the spinor invariant.
Then the transformation \nref{sltwo} will deform the theory that lives at singularity.
This same transformation can give us the near horizon geometry of the new theory.

There is no reason we should restrict to conformal field theories, we can do
this type of deformation in non-conformal field theories. The simplest examples
would be the theories living on D-p branes in flat space \cite{sunny}. More complicated
examples include the cascading theory considered by Klebanov-Strassler
\cite{ikms}.

So far we have been discussing the theory on the brane that arises after putting
branes in backgrounds where we have performed our $SL(2,R)$ transformation. We can
as easily consider the gravity duals of these theories. If we know the
gravity dual of the field theory living on a D-brane in the original background,
then the gravity dual of the deformed field theory living on the D-brane on
the new background is given by performing the $SL(2,R)$ transformation
\nref{sltwo} on the original solution.
We discuss explicit examples in the following sections.

What we have discussed so far applies both for IIA and IIB string theory. In
fact, some of our discussion also applies for bosonic string theory.
Let us now concentrate in the case of IIB supergravity. This theory has
an $SL(2,R)_s$ symmetry already in ten dimensions\footnote{The subindex $s$ in
$SL(2,R)_s$ was introduced to avoid confusion with other $SL(2,R)$ symmetries
of the problem.}. Once we compactify the
theory on a two torus, this $SL(2,R)_s$, together with the $SL(2,R)$ symmetry
that acts on \nref{taudef} form an $SL(3,R)$ group.
These can be used to generate more general solutions. Perhaps a simple way to start
is to first
   do an S duality, then
the transformation \nref{sltwo}, and finally an S-duality again. This combined
transformation acts on a tau parameter that involves the RR field.
Actually, a compactification of
type IIB supergravity on a two torus has $SL(3,R)\times SL(2,R)$ symmetry
\footnote{The $SL(3,R)$ symmetry is most clearly seen by viewing
this as a compactification of M-theory on $T^3$.}.
The second $SL(2,R)$ is the one acting on the complex structure of the two torus and
will not play an important role now.
Starting with a general non-singular ten dimensional  solution of type IIB supergravity
with two $U(1)$ geometric symmetries we can act with $SL(3,R)$ transformations to
generate new solutions. If we are interested in generating non-singular solutions,
starting from solutions where the two torus goes to zero in some regions, then
the $SL(3)$ matrix cannot be arbitrary. It to be of the form
 \be \label{slthree}
h_3 = \left( \begin{array}{ccc} 1 &0 & 0  \\ \gamma & a & b \\
\sigma & c & d \end{array} \right) ~,~~~~~{\rm where} ~~~~~~
h_2 \equiv \left( \begin{array}{cc}  a & b \\
  c & d \end{array} \right) ~\in  SL(2,R)
\ee
In total, such solutions contain four  parameters.  Two of them arise
 from  the axion-dilaton
$\tau_s$
which parametrizes $SL(2,R)/U(1)$ associated to the matrix $h_2$ \footnote{
Do not confuse $\tau_s$, which parametrizes the eight dimensional axion dilaton,
with $\tau$ defined in \nref{taudef}.}. The
other two parameters
are $\gamma$ and $\sigma$. The latter are periodic variables
$\gamma \sim \gamma+1, ~\sigma \sim \sigma + 1$ and they form a representation
of $SL(2,Z)_s$ which acts on $\tau_s$.

If we have more $U(1)$ symmetries we can do other
deformations. For example, let us consider an eleven dimensional supergravity solution
with three $U(1)$ symmetries. This solution contains a three torus. The dimensional
reduction of eleven dimensional supergravity
on a $T^3$ is the same as IIB on the $T^2$ and   has an $SL(3,R)\times SL(2,R)$.
We can do an $SL(2,R)$ transformation on
\be
\tau = C_{123} + i \sqrt{g}
\ee
where $\sqrt{g}$ is the volume of the three torus.
We discuss this a bit further in section 5.

\section{ Marginal deformations of ${\cal N} =4$ Super Yang Mills}
\renewcommand{\theequation}{3.\arabic{equation}}
\setcounter{equation}{0}

\subsection{Field theory}

As studied in \cite{marginal}\cite{rlms}, there is a three parameter family of
marginal deformations of ${\cal N}=4$ Super Yang Mills that preserve ${\cal N}=1$
supersymmetry.
These theories have a superpotential of
the form
\be \la{superpotls}
h \, Tr( e^{ i  \pi \beta} \Phi_1\Phi_2 \Phi_3 - e^{- i  \pi \beta}
 \Phi_1\Phi_3 \Phi_2  ) + h' Tr( \Phi_1^3 +\Phi_2^3 +\Phi_3^3 )
\ee
where $\Phi_i$ are the three chiral superfields.  The parameters
$h, h' , \beta$ are complex. In addition we have the gauge coupling. The
condition of conformal invariance imposes only one condition among these four parameters
\cite{rlms}. In this paper we set $h'=0$. In this case, besides the $U(1)_R$
symmetry,  we have
a  $U(1) \times U(1)$ global symmetry generated by
\bea \nonumber
U(1)_1:& ~~~~&(\Phi_1,\Phi_2 ,\Phi_3 ) \to (\Phi_1,e^{i\varphi_1}\Phi_2 ,e^{-i \varphi_1} \Phi_3 )
\\\la{twouone}
U(1)_2:&~~~~&(\Phi_1,\Phi_2 ,\Phi_3 ) \to (e^{-i \varphi_2} \Phi_1,e^{i\varphi_2}\Phi_2 , \Phi_3 )
\eea
This symmetry leaves the superpotential invariant. It also leaves the supercharges
invariant.
The
Leigh-Strassler argument says that
 we have a 2 dimensional manifold of ${\cal N} =1$ CFTs with $U(1)\times U(1)$ global
symmetry  (beyond the usual $U(1)_R$ symmetry).
This field theory been studied previously in
\cite{berenstein1,berenstein2,berenstein3,ofer,doreysduality,doreynsfive,doreycoulomb,benini,hollowood,niarchos}.
Note that if we start with  $U(N)$ theory the fields $Tr[\Phi_i]$ couple through the
superpotential \nref{superpotls} to the $SU(N)$ fields. It was noted in \cite{hollowood}
that these couplings flow to zero in the IR. So in the IR we have an $SU(N)$ theory.

We will now review a few aspects of this field theory. The theory is
invariant under a discrete $Z_3$ symmetry which acts as a
cyclic permutation of the three chiral fields.
The physics turns out to be periodic in the variable $\beta$. In fact, we can think
of the variable $\beta$ as living on a torus with complex structure $\tau_s$, where
$\tau_s $ is related to  gauge coupling and theta parameter
of the field theory \cite{doreysduality}.
The theory has an $SL(2,Z)$ duality group. The variable $\tau_s$ was chosen in
\cite{doreysduality} so that it transforms in the usual way under S-duality.
Then $\beta$ transforms as a modular form \cite{doreysduality}
 (see also \cite{intriligator}).
In other words
\be \la{sduality}
\tau_s \to { a \tau_s + b \over c \tau_s + d} ~,~~~~~~~~~\beta \to { \beta
 \over c\tau_s + d} ~,~~~~~~~~~~ \left( \begin{array}{cc}  a & b \\
  c & d \end{array} \right) ~\in  SL(2,Z)
\ee
The periodicity in $\beta$ is
\be
\beta \sim \beta + 1 \sim \beta + \tau_s
\ee
The first identification is obvious from the superpotential \nref{superpotls}. The
second is not clear perturbatively, but it follows from S-duality.
It is also useful to parametrize $\beta$ as
\be \la{gamsigdef}
\beta = \gamma - \tau_s \sigma
\ee
where $\gamma$ and $\sigma $ are real variables with period one.
Note that the description of the $\beta$ deformation as a star product, as in \nref{deform}
is valid for real $\beta$. For complex $\beta$ we are just complexifying this parameter
in the superpotential, which is not the same as using the prescription
\nref{deform} with complex $\beta$ to construct the component lagrangian.

Let us find the Coulomb branch. A general analysis of the Coulomb branch of this
theory can be found in \cite{doreycoulomb,benini}.
The F-term constraints can be written as
\be \label{nctorus}
\Phi_2 \Phi_3 = q \Phi_3\Phi_2 ~,~~~~~~~~
\Phi_1 \Phi_2 = q \Phi_2\Phi_1 ~,~~~~~~~~
\Phi_3 \Phi_1 = q \Phi_1\Phi_3  ~,~~~~~~~~q = e^{ 2 \pi  i \beta}
\ee
For generic $\beta$
the Coulomb branch consists of diagonal matrices where in each entry two of the $\Phi_i$
are zero and only one is non-zero. So we can have, for example $\Phi_1 \not =0$ and
$\Phi_2=\Phi_2=0$.
For $\sigma =0$ and a rational $\gamma$ we have new features,
there are additional regions in the Coulomb branch where the matrices form the fuzzy
torus algebra. For $\gamma = m/n$ we can solve \nref{nctorus} with \cite{berenstein1}
\be \la{coulbr}
\Phi_1 = x U^m ~,~~~~~~ \Phi_2= y V~,~~~~~~~~ \Phi_3 = z V^{-1} U^{-m} ~,~~~~~~~~
 UV=e^{ i 2 \pi/n} VU
\ee
with $U$ and $V$ the standard matrices of the non-commutative torus\footnote{
$U = diag(1,e^{i 2\pi/n}, \cdots, e^{ i 2\pi (n-1)/n} )$, and $V$ has non-vanishing
elements $V_{i+1,i} =   V_{1,n}=1$.} and
$x,~y,~z$ are complex numbers.
We can think of these solutions as D3 branes forming toroidal D5 branes via the
Myers effect \cite{myers}.

S-duality implies that there should be such additional branches whenever
the complex variable $\beta$ is a rational point on the torus. In other words,
whenever there is an integer $n$ such that the point $n \beta$ is a point on the
lattice generated by $1, \tau_s $. If
\be \la{ratorus}
n (\gamma, \sigma)  = (p , q)~,~~~~~~~~~~n,p,q \in Z
\ee
we expect that $n$ D3 branes can form a toroidal $(p,q)$ fivebrane. The appearance of
this branch is not clear from perturbation theory, but was explored via matrix model
techniques \cite{rdcv} in \cite{doreycoulomb,benini}.
Of course, we can only have these
branches in theories where $N \geq n$.

In addition, the $SU(N)$ theory  has another  branch \footnote{This
can be obtained from a massless limit of the one discussed in  \cite{doreysduality}.}, even
for generic $\beta$.
This branch arises because
in $SU(N)$ we have to project the equations \nref{nctorus} into their traceless part.
So we can have solutions where, for example, the matrices $\Phi_i$ are diagonal and
such that $(\Phi_i)_{ll} (\Phi_j)_{ll} $ (no sum) is a constant independent of the
matrix index $l$. In this paper, we will not  describe this branch from
the supergravity side.

It is also interesting to consider the chiral primary operators since they correspond
to
  BPS states on the gravity side.
For generic $\beta$ there is a single single trace BPS operator
with each of the following charges \cite{berenstein1,berenstein2}
\be \la{chgen}
(J_1,J_2,J_3) = (k,0,0), ~~~(0,k,0), ~~~~(0,0,k) ~,~~~~~(k,k,k)
\ee
where $k$ is an arbitrary integer.
 In addition we can have multitrace operators formed  by products of
single trace operators.    For $\sigma =0$ and $\gamma= m/n$
($m,n$ coprime) there are
more single trace operators with
\be  \la{modn}
 (J_1,J_2,J_3) = (k_1,k_2,k_3) ~,~~~~~~~~k_1=k_2=k_3 ~{\rm mod}(n)
\ee
We expect that the chiral ring has special features for more
generic rational $\beta$ \nref{ratorus}, but we are not aware of a discussion
of this point in the literature.

\subsection{Supergravity solution}

It follows from the analysis of the Kaluza Klein spectrum on $AdS_5\times S^5$
that there is a massless field in $AdS_5$ that corresponds to the deformation in
question. In fact, there are more massless fields in  $AdS_5$ than there are
exactly marginal deformations.  In \cite{kol} the super-gravity equations were analyzed
to second order and a constraint was found. There are as many solutions to this
constraint as there are exactly marginal deformations of ${\cal N}=4$.\footnote{
Singular solutions were found in \cite{ansar}.  They are not the gravity duals
of the $\beta$-deformations, which are non-singular. }

In this section we will show how to get the exact solution   for
deformations which preserve $U(1)\times U(1)$ global symmetry. All we need to do
is to apply the method described in the previous section.
Let us start by writing the metric of $S^5$ in the form\footnote{
To save space we define $s_\alpha = \sin \alpha$, $c_\alpha = \cos \alpha$, etc. }
\bea
{ ds^2 \over R^2} &= & \sum_{i=1}^3  d\mu_i^2 + \mu_i^2 d\phi_i^2 ~,~~~~~~~~~~~
{\rm with}~~~ \sum_i \mu_i^2 =1
\\
{ ds^2 \over R^2} &=&
 d\alpha^2 + s_{\alpha}^2 d\theta^2 + c_\alpha^2(d\psi - d\varphi_2)^2
+ s_\alpha^2 c_\theta^2 (d\psi + d\varphi_1 + d\varphi_2)^2 +
s_\alpha^2 s_\theta^2 (d\psi - d\varphi_1 )^2
\nonumber\\
&= & d\alpha^2 + s_{\alpha}^2 d\theta^2+
{ 9 c_\alpha^2 s^2_\alpha s^2_{2 \theta} \over 4 c_\alpha^2 +
s_\alpha^2 s^2_{2 \theta} }d\psi^2 +
\la{laslin}\\
&& + s^2_\alpha \left[ d\varphi_1 + c^2_\theta d\varphi_2 +
  c_{2\theta} d\psi\right]^2+ (c_\alpha^2 + s_\alpha^2 s^2_{\theta}c^2_\theta )
\left[ d\varphi_2 + { (- c_\alpha^2 + 2 s^2_\alpha s^2_{  \theta}c^2_\theta)
\over   c_\alpha^2 + s_\alpha^2 s^2_{  \theta} c^2_\theta} d\psi \right]^2
   \nonumber
\eea
Notice that the two $U(1)$ symmetries \nref{twouone} act by shifting
$\varphi_1,~\varphi_2$. So the two torus we were talking about in
our general discussion has a metric given by the last line in \nref{laslin}.
Actually, we can compute the $\tau$ parameter of this two torus
\be \la{tauads}
\tau = i \sqrt{g_0} = i [R^2 s^2_\alpha (c^2_\alpha + s_\alpha^2 s_\theta^2 c_\theta^2)]^{1/2} 
=i R (\mu_1^2 \mu_2^2 + \mu_2^2 \mu_3^2 + \mu_1^2 \mu_3^2)^{1/2}
\ee
where $R = (4\pi g_s N)^{1/4}$.
Note that the terms involving $\psi$ in the last line \nref{laslin}
become gauge fields in the eight dimensional theory.
After we apply the transformation \nref{sltwo} we can find the solution
corresponding to the gravity dual of the deformed theory
\bea\label{GammaDeform}
ds^2_{str} &=& R^2 \left[ ds^2_{AdS_5} +  \sum_i ( d\mu_i^2  + G \mu_i^2 d\phi_i^2) + \hat \gamma^2
G \mu_1^2\mu_2^2\mu_3^2 (\sum_i d\phi_i)^2 \right] \la{metrgam}
\\
G^{-1} &=& 1 + \hat \gamma^2 (\mu_1^2 \mu_2^2 + \mu_2^2 \mu_3^2 + \mu_1^2 \mu_3^2) ~,~~~~~
\hat \gamma = R^2 \gamma  \la{defofggam}  ~,~~~~~~~R^4 \equiv 4 \pi e^{\phi_0} N
\\
e^{2 \phi} &=& e^{2 \phi_0} G\nonumber
\\
B^{NS} &=&  \hat \gamma R^2G ( \mu_1^2 \mu_2^2 d\phi_1 d\phi_2 +
\mu_2^2 \mu_3^2 d\phi_2 d\phi_3 +
 \mu_3^2 \mu_1^2 d\phi_3 d\phi_1 )\nonumber
 \\
 C_2  &=& -3\gamma(16 \pi  N) w_1 d\psi ~,~~~~~~~~~~~~~{\rm with}~~~ d w_1 =  c_\alpha s_\alpha^3
 s_{\theta} c_\theta d\alpha d\theta\nonumber
 \\
 C_4  &=& (16 \pi N) ( w_4 + G w_1 d\phi_1 d\phi_2 d\phi_3 )\nonumber
 \\
 F_5 &=& (16 \pi N) ( \omega_{AdS_5} +  G \omega_{S^5}) ~,~~~~~~~~~~
 \omega_{S^5} = dw_1 d\phi_1 d\phi_2 d\phi_3 ~,~~~~~\omega_{AdS_5} = dw_4\nonumber
 \eea
 where $\omega_{S^5}$ is the volume element of a unit radius $S^5$.
 The metric is written in string frame.
  We have made manifest the $Z_3$ symmetry
   which is broken in the intermediate stages when
 we write the metric as in \nref{laslin}.

Let us first examine the regime of validity of this solution. The solution which is
presented here has small curvature as long as
\be \la{curv}
 R \gamma  \ll 1   ~,~~~~~R \gg 1
 \ee
The first inequality can be understood as the condition that (at a generic point)
the two torus does not
become smaller than the string scale after the transformation.
We also computed the square of the Riemann tensor on
the deformed fivesphere and  looked at the region where it is a maximum as a check
of the first condition (\ref{curv}).

Suppose that  $\gamma =1/n$. For these cases the general argument
presented in \nref{taudp} shows that the solution is T-dual (or, more precisely,
$SL(2,Z)$ equivalent) to a $Z_n\times Z_n$ orbifold with discrete torsion
\cite{douglasfiol,berenstein1}.
Then we see from \nref{curv} that the solution presented above \nref{metrgam}
 has high curvature
 if $n$ is a relatively small number, while
the orbifold description will be weakly coupled.  On the other
hand, if
\be \la{ncond}
 n \gg R \sim (g^2_{YM} N)^{1\over 4}
 \ee
 our
  solution will be weakly curved. But in this case the orbifold description would
  contain circles smaller than the string scale and would not be a good description.
If wanted to check that $\gamma$ is a periodic variable, $\gamma \sim
\gamma +1$,  we have
to go trough a region where we do not trust the gravity solution. Nevertheless, by
construction,  our solution
formally has this periodicity, after doing the appropriate duality ($SL(2,Z)$
 transformation).
 The topology of this solution \nref{metrgam} is always $AdS_5\times S^5$, since our
 transformation \nref{sltwo} does not change the topology.

\subsection{Coulomb branches and chiral primaries}

We now explore the special Coulomb branches that appear for rational $\gamma$.
Perhaps a simple way to understand the emergence of these branches is the following.
The eight dimensional theory that we obtain after compactification on the two torus
has an $SL(3)$ symmetry. The original D3 branes, together with the D5 and NS5 wrapped
on the two torus transform under the fundamental  representation of $SL(3)$. In
other words we have the transformation law
\be
 (N_{D3}, N_{D5} , N_{NS5}) \to (N_{D3} , N_{D5} + \gamma N_{D3} , N_{NS5}  )
\ee
under \nref{sltwo}. For non-zero $\sigma$ we will also shift the NS5-brane charge.
So, suppose that we start from some number of D3 branes in the
Coulomb branch, at some point where the two torus has finite size. Then,
after applying \nref{sltwo} we see
that the D5 charge will be proportional to $\gamma N_{D3}$.
In general this does not obey the D5 charge quantization condition and is not
an allowed brane. But for
the special values $\gamma = m/n$ and $N_{D3}$ a multiple of $n$, then we do obey
the D5 charge quantization condition. Since the original configuration was BPS, this
new configuration will also be BPS.

Just to check our formulas, let us
look at this  branch more explicitly in the probe approximation. Notice
that by taking first the limit $N\to \infty$, keeping $n$ and $R$ fixed such that
\nref{curv} \nref{ncond} are
 obeyed, we can ensure that the probe approximation is valid.
Let us  write down the Dirac-Born-Infeld action for the D5 branes, and let us set
the RR scalar to zero for simplicity.
\be
S = \int  e^{-\phi} \sqrt{ det (g+ F - B) } - \int C_6 + (F - B)  \wedge C_4
\ee
We will now show that $C_6 - B\wedge  C_4$ is zero.
$C_6$ is determined by the equation\footnote{ See
\cite{jpms}, for example. To write these equations collect all RR fields into
a single form $C = C_0 + C_2 + \cdots$. Then the field strength
$G = d C - H \wedge C $ is gauge invariant under $C\to C + d\Lambda - H \wedge \Lambda$
and self dual $G= * G$. The gauge invariant coupling to D-branes is $\int e^{F-B} C $.}
\be \la{csix}
 dC_6 =  *dC_2 +  H_3  \wedge C_4 ~,~~~~~~H = d B
 \ee
 So we find that the particular combination coupling to the D5 brane obeys
 the equation
 \be \la{eomrr}
 d (C_6 -  B  \wedge C_4) =  *dC_2  - B  \wedge d C_4 =
  *dC_2  - B \wedge F_5
  \ee
where we used that in our background $(B )^2 dC_2 =0$. It can be checked
that the two terms in the right hand side of \nref{eomrr} cancel each other.
\footnote{
 The equations of motion for $ C_2$ ensure that the exterior derivative of
 the right hand side is zero. It turns out  that the two terms
 in the right hand side of \nref{eomrr} have the same functional form, so that
 they could differ just by a numerical coefficient. It is clear that the numerical
 coefficients should be such that the two last terms in \nref{eomrr} cancel
 precisely. Otherwise, its exterior derivative would be non-zero.}
 Since \nref{eomrr}
 is zero,  the coupling of the 5-brane to $C_6 - B \wedge C_4$
 is constant. We can go to a region where the D5 shrinks to a D3 brane where this
 coupling is absent. So  we conclude that the constant in question is zero.
Thus, we find that the DBI action for a D5 brane wrapping the two torus and
extended along a Poincare slice of $AdS_5$ is
\be
S \sim  r^4 N_5 \left[
e^{-\phi} \sqrt{ G^2 g_0  + ( F_{12} - B_{12})^2 }  - F_{12} \right]
 \la{actdfive} \ee
where the factor of $r^4$ comes from the $AdS$ part, $N_5$ is the number of D5 branes,
$g_0$ is the determinant of the metric
of the original two torus \nref{tauads} and $G$ is the function in \nref{defofggam}.
If we pick
$F_{12} = 1/\gamma $, then  the action \nref{actdfive}
vanishes, due to the form of the $B$ field and dilaton  in  \nref{sltwo}.
Of course, in order to obey the quantization condition for D3 brane
charge (or the $U(1)$ flux on the D5 brane)
we need that $N_5 \gamma^{-1}$ is an integer. This can be obeyed if
$\gamma = m/n$ and $N_5=m$. This branch corresponds to the non-commutative branch
described in \nref{coulbr}.

Suppose that we sit at a point in the Coulomb branch with a large number of
coincident D5 branes so that we cannot ignore their backreaction on the geometry.
We can find the resulting gravity solution in the following way\footnote{
These metrics were described in an approximate way
in \cite{doreynsfive}.}. We start from a solution dual to
${\cal N}=4$ SYM on the Coulomb branch. We choose this solution so that it
contains D3 branes smeared on the two torus appearing in our discussion. Then
we perform the $SL(2,R)$ transformation \nref{sltwo}. This gives us
a solution which is regular everywhere except on the two-tori where we smeared the
original D3 branes. In the vicinity of the two torus the metric looks similar to
that of D5 branes on a non-commutative two torus \footnote{
In fact, the gravity solutions for non-commutative field theories  \cite{sunnyh,jmjr}
can be obtained
by applying the $SL(2,R)$ transformation \nref{sltwo}
to two of the worldvolume directions of
the gravity solution dual to the ordinary field theory.}.
 But extremely close to the two torus, one
would need to perform T-dualities, similar to those described in \cite{sunny} which
would give us $AdS_5\times S^5$ in the extreme IR. The S-dual of this
configuration was
explored in \cite{doreynsfive} as a deconstruction of the NS5 brane theory on a
two torus.

Let us analyze the chiral primary states.
Let us first view the chiral primaries as classical string solutions. This will
be a good description when their charges  are large.
Then we see that pointlike strings, or particles,  with momenta
\be
(J_1,J_2,J_3) = (J,0,0),~~~(0,J,0),~~~~(0,0,J),~~~~(J,J,J)
\ee
lead to BPS states for any $\gamma$. This agrees with the field theory analysis.

Something more interesting happens at $\gamma =m/n$. In this case we can
get BPS solutions which contain strings wrapped on the cycles of the two torus
on which we are doing the $SL(2,R)$ transformation. These are contractible cycles in
the full geometry, so this winding is topologically trivial. Nevertheless, this
implies that the corresponding string states are macroscopic strings and are not pointlike.
The resulting string states are rather similar to those analyzed by
\cite{frolov}. Of course these are solutions to the equations
of motion for the string. Again, the easiest way to find the solutions is to formally
do the $SL(2,R)$ transformation starting from a
generic BPS pointlike string in the original background. This state will have
momenta $n_1,n_2$ on the two circles of the two torus. Formally, after the transformation
\nref{sltwo}
the momenta are the same and the winding numbers are $w_1 = -\gamma n_2 ,~~
w_2 =  \gamma n_1$
Again, when $\gamma =m/n$   the quantization for winding is obeyed if $n_1,~n_2$ are
multiples of $n$. This condition has to be obeyed if we are considering
a classical string solution. The fact that $n_1,n_2$ are multiples of $n$ implies
that $J_1=J_2=J_3$ mod$(n)$, as in \nref{modn}. In appendix B we present the explicit
classical solutions.

It should be noted that some of the general solutions for BPS states presented in
\cite{llm} can be used to generate similar solutions in our case.
Namely, if we start from a solution in \cite{llm} representing a half BPS state
that corresponds to a fermion droplet that is circularly symmetric, then the ten
dimensional solution will have the two necessary isometries so that we can
apply our $SL(2,R)$ transformation \nref{sltwo}. In this way we generate
gravity solutions corresponding to some of the BPS states of the $\beta$-deformed
theory.

\subsection{General deformation and S-duality}

So far we have been discussing the case when $\sigma =0$, we can easily generate
the solutions for non-zero $\sigma$ by performing $SL(2,R)_s$ transformations
of the solutions with $\sigma =0$. Here we are referring to the $SL(2,R)_s$
symmetry group of the ten dimensional theory, which should not be confused with
the $SL(2,R)$ group that we used in \nref{sltwo}. In other words, start with
$AdS_5 \times S^5$ and we perform
a more general $SL(3,R)$ transformation in the eight dimensional theory.
The resulting solution is the following.
\bea
ds^2_E &=& R_E^2 G^{-1/4} \left[ ds^2_{AdS_5} +
   \sum_i ( d\mu_i^2  + G \mu_i^2 d\phi_i^2) +{ | \gamma -\tau_s \sigma|^2 \over \tau_{2 s} }
R_E^4 \mu_1^2\mu_2^2\mu_3^2 (\sum_i d\phi_i)^2 \right] \la{metrgen}
\\
e^{-\phi} &=&    { \tau_{2 s}} G^{-1/2} H^{-1} ~,~~~~~~~~~~~\chi =  { \tau_{2 s} } \sigma
(\gamma - \tau_{1 s} \sigma) H^{-1} g_{0,E} + \tau_{1 s}
\\\la{defofggen}
G^{-1} &\equiv& 1 + { | \gamma -\tau_s \sigma|^2 \over \tau_{2 s} } g_{0,E} ~,~~~~~~~~~~~
H   \equiv   1 +
 { \tau_{2 s} \sigma^2 } g_{0,E} ~,~~~~\tau_s = \tau_{1s} + i \tau_{2s}
 \\ \la{gzeroe}
 g_{0,E} &=& R_E^4
 (\mu_1^2 \mu_2^2 + \mu_2^2 \mu_3^2 + \mu_1^2 \mu_3^2) ~,~~~~~
R_E^4 = 4 \pi N
 \\
B^{NS} &=&  { \gamma -\tau_{1 s} \sigma  \over \tau_{ 2s } }R_E^4 G
 w_2  -     \sigma 12 R_E^4 w_1 d\psi
 \\
 \nonumber
 C_2  &=&
 [ - \tau_{2 s} \sigma + { \tau_{1 s} \over \tau_{2 s} } (\gamma -
 \tau_{1 s} \sigma ) ]  R_E^4 G w_2
-  \gamma 12 R_E^4 w_1 d\psi
\\ \nonumber
  &~&   d w_1 =  c_\alpha s_\alpha^3 s_{\theta} c_{\theta} d\alpha d\theta
  ~,~~~~~~~w_2 = ( \mu_1^2 \mu_2^2 d\phi_1 d\phi_2 +
\mu_2^2 \mu_3^2 d\phi_2 d\phi_3 +
 \mu_3^2 \mu_1^2 d\phi_3 d\phi_1 )
 \\
 F_5 &=& 4 R_E^4 ( \omega_{AdS_5} +  G \omega_{S^5}) ~,~~~~~~~\omega_{S^5} = dw_1 d\phi_1
 d\phi_2 d\phi_3
 \eea
where we wrote the Einstein metric.
This metric depends on the complex parameter $\tau_s$, which we identify
with the $\tau_s$ of the field theory and also on $\gamma , \sigma$ which we
identify with the same parameters of the gauge theory \nref{gamsigdef}. We can
check that the solution transforms appropriately under S-duality transformations
\nref{sduality}.

An interesting question to ask is: What is the metric in the space of couplings?.
In other words, what is the Zamolodchikov metric for these marginal deformations.
In the supergravity regime this metric can be computed from our solution.
It is simplest to think about this problem from the eight
dimensional point of view, after reducing on the two torus. The eight dimensional
field theory has an $SL(3,R)$ invariance (plus another
 $SL(2,R)$ symmetry that is not important
now) and it has scalar fields
living on the $SL(3,R)/SO(3)$ coset. The fields $\tau_s , \gamma , \sigma$ parametrize
some of the fields in this coset. The  fifth field is a scalar that measures the size
of the two torus in the IIB language. We are not free to change it if we want to
have a solution that is non-singular in ten dimensions. $SL(3,R)$
invariance then determines
the action in eight dimensions, which in turn gives us the five dimensional action on
$AdS_5$ for
  the massless scalar fields in $AdS_5$
which are dual to the marginal deformations.
The bulk action is
\be \la{zzmet}
S =  { N^2 \over 16 \pi^2} \int_{AdS_5}   \left[  { \partial \tau_s \partial \bar \tau_s\over \tau_{2 \, s}^2 } +
C
 { | \partial \gamma -
 \tau_s \partial \sigma |^2 \over \tau_{2 \, s} } \right]
\ee
where the integral is over an $AdS_5$ space of radius one and $C$ is a constant.
The overall coefficient is the same as the one appearing in the computation
in \cite{sgik}, which determines the coefficient for the dilaton.
This result \nref{zzmet} is derived as follows.
We first write the action in eight dimensions.
  It is
convenient to parametrize an element of $SL(3)/SO(3)$ as
\be \la{parcos}
l =  \left( \begin{array}{ccc} 1 & 0 & 0 \\ \gamma & \sqrt{\tau_{2 s}} & { \tau_{1 s}
\over \sqrt{\tau_{2 \, s} } }
\\ \sigma & 0 & {1\over \sqrt{\tau_{2 s}} } \end{array} \right)
 \left( \begin{array}{ccc} g^{1/3} & 0 & 0 \\ 0 & g^{-1/6} & 0 \\
 0 & 0 & g^{-1/6}   \end{array} \right)
\ee
Here $g$ is the determinant of the metric on the internal two torus. We can
then compute the metric by computing the matrix $M = l l^t$ which contains only the
gauge invariant information on the coset. The parametrization \nref{parcos} amounts
to a choice of gauge. The metric in the gravity solution is determined as
\be
 Tr[ \partial M \partial M^{-1} ]
 \ee
 We see that this metric is $SL(3,R)$ invariant. Computing it for \nref{parcos} we find
a metric like \nref{zzmet} were $C \to  g_{0,E}$ where $g_{0,E}$
is defined in \nref{gzeroe} and is a function of some of the angles of the sphere.
Finally we conclude that $C$ in \nref{zzmet} is
\be \la{valofc}
C = \langle g_{0,E} \rangle = R_E^4 \langle \mu_1^2 \mu_2^2 + \mu_2^2 \mu_3^2 + \mu_1^2 \mu_3^2
\rangle =
 \pi N
\ee
where the average is over the five sphere. This average arises because we are interested
in the five dimensional action on $AdS_5$. We see that the $N$ and $g_{YM}$ dependence
match the weak coupling result. We have not checked explicitly the overall numerical
coefficient \footnote{
Note that it makes sense to compare the normalization of the two point functions because
  these are marginal operators and the parameter $\gamma$ has a natural normalization,
where it has period $\gamma \sim \gamma +1$. }, but the results of
\cite{vanigor} imply that the weak coupling
result will be the same as the gravity result.
The authors of  \cite{vanigor} considered
two point functions
of operators associated to higher Kaluza-Klein harmonics of the dilaton. The first
Kaluza-Klein harmonic of the dilaton is
 related by ${\cal N}=4$ supersymmetry to   the
operator corresponding to a small change in $\gamma $.

The discussion about the Coulomb branch is very similar to what we had before.
Whenever we are at a rational point on the two torus \nref{ratorus} we can have
a $(p,q)$ fivebrane in the bulk describing a new piece of the Coulomb branch.
Similarly, there are BPS chiral primary states associated to  $(p,q)$ strings
wrapping the two torus.

\subsection{pp-wave limit}

Starting from the general solution \nref{metrgen} we can take interesting  pp-wave
limits. For example, we can start with a maximum circle on the deformed $S^5$ at $\mu_1=1$.
Lightlike trajectories along this circle, which remain at the origin in $AdS_5$,
correspond to BPS operators of the form $Tr[\Phi_1^{J}]$.
We can now take a pp-wave limit
\bea \label{pplim}
 &&J, ~R \to  \infty,\quad {\rm with} \quad -p_- =  { J \over R^2}
 ~,~~ ~~\tilde \gamma \equiv
{ (\gamma - \tau_{1 \, s} \sigma)   R^2 },\quad
\tilde \sigma \equiv \tau_{2 \, s}{ \sigma   R^2}\quad{\rm fixed}
\nonumber\\
&&\qquad x^-  =  (t - \phi^1 ) R^2,\quad x^+ = t ~,~~~~{\rm fixed}
\eea
with the rest of the coordinates in $AdS_5$ and (the deformed)  $S^5$ scaling in the same
was as the usual pp-wave limit \cite{blau,bmn}.

The metric then becomes
\bea \la{ppwave}
ds^2 &=& - 2 dx^+ dx^- - [r^2 + (1 + \tilde \gamma^2 + \tilde \sigma^2)y^2](dx^+)^2 +
d\vec y^{ \, 2}  + d \vec r^{ \, 2}
\\
H^{NS} &=& - 2 \tilde \gamma dx^+ (dy_1 dy_2 - dy_3 dy_4)
\\
H^{RR} - \tau_{1 \, s} H^{NS} & = & 2 \tau_{2 \, s} \tilde \sigma dx^+ (dy_1 dy_2 - dy_3 dy_4)
\\
F_5 & =& {\tau_{2 \, s} }dx^+ (dr_1dr_2dr_3dr_4 + dy_1dy_2 dy_3 dy_4)
\eea
For simplicity let us set the axion to zero $\tau_{1 \, s} =0$.
When we quantize strings in lightcone gauge with $x^+ =  \tau$ we find a massive theory
on the string worldsheet.
The oscillators in the $r$ directions
have the same spectrum as in the standard $AdS$ case, \cite{bmn}. The oscillators in
the $y$ directions have the spectrum
\be \la{sepctr}
\omega_{n \, y \pm } = \sqrt{ 1 +   \left( { n \over  |p_-| } \pm \tilde \gamma \right)^2    + \tilde \sigma^2 } =
\sqrt{ 1 + ( 4 \pi g_s N) |{ n \over J} \pm \beta |^2
    }
\ee
where the $\pm$ indicates the spin on the $y^1, y^2$ or $y^2, y^4$ planes.
In fact, this pp-wave and its relation to the BMN limit of the field theory were
studied by Niarchos and Prezas \cite{niarchos}. They started from the field
theory and they  derived the result \nref{sepctr} by generalizing the arguments in
\cite{santambrogio}. Then they realized that these results could be reproduced by
the pp-wave in \nref{ppwave}. Here we have shown how \nref{ppwave} arises from the
full ten dimensional geometry.
In general, on the pp-wave we
see what we have BPS states whenever $\sigma=0$ and $ n = |p_-| \tilde \gamma$.
In terms of the original variables we could write
$\beta = \gamma =1/n'$,
$J = n' k$ and $n=k$ where $k$ is not necessarily large in the limit \nref{pplim}.

We could also take a different pp-wave limit by looking at states having charges
near to $ (J_1,J_2,J_3) \sim (J,J,J) $. In this case  we obtain a different pp-wave.
This pp-wave can be obtained by taking the appropriate limit of our metric or by
performing an $SL(2,R)$ transformation on the maximally supersymmetric pp-wave
in ``magnetic'' coordinates.
Namely, we start with \cite{pando}
\be \la{otherpp}
ds^2 = - dx^+ dx^-  - r^2 (dx^+)^2  + d{\vec r}^{ \, 2} +
4 dx^+( y_1 dy_2 + y_3 dy_4) + d{\vec y}^{\, 2}
\ee
and then we apply the $SL(2,R)$ transformation \nref{sltwo} by considering
a two torus parametrized by $y^2$, $y^4$. These coordinates are non-compact, but
we can compactify them, perform \nref{sltwo}, and then decompactify them again.
The resulting metric is the pp-wave limit of the  solution \nref{metrgam} along
the null geodesic $\psi = t$.
We have not checked whether results computed using this new pp-wave limit agree with
expectations from the gauge theory side.

It is natural to ask if the $\beta$-deformed  field theory leads to
an integrable spin chain. In \cite{roiban}\cite{cherkis} this question was
answered in the affirmative in the sector made with chiral fields
$\Phi_1,\Phi_2$ for generic $\beta$ or in the sector with all
three chiral fields $\Phi_1,\Phi_2,\Phi_3$ with $\beta$ real,
while is it not integrable for complex $\beta$. These results are
important for the study of the pp-wave resulting from an $SL(2,R)$
transformation of \nref{otherpp}.

\section{ More general theories}
\renewcommand{\theequation}{4.\arabic{equation}}
\setcounter{equation}{0}

In this section we extend the previous discussion to more general
theories. The simplest examples are orbifolds of $AdS_ 5 \times
S^5$. It is clear from our construction of the marginal
deformation as a star product that the resulting theories are
conformal to leading order in $N$, since planar diagrams are the same as in the
original   theory \cite{filk}. In the case that we have a
$Z_n$ orbifold that preserves at least ${\cal N}=1$ supersymmetry,
then the results of \cite{ofer,razamat} show that the marginal
deformations of the type we are considering are exactly marginal
and preserve ${\cal N}=1$ superconformal invariance. In these
cases the $Z_n$ orbifold action is a subgroup of the $U(1)\times
U(1)$ non-R-symmetry that was used to construct the marginal
deformation \nref{deform}. The metrics for these deformations are
just simply   orbifolds of \nref{metrgen}.

  A conformal field theory that is particularly interesting
  is
the theory living on D3 branes on a conifold \cite{ikew}, see
\cite{conifoldreview} for a review. This theory has an
$SU(2)\times SU(2)$ global symmetry that leaves the supercharges
invariant. We can consider a $U(1) \times U(1)$ subgroup of $SU(2)
\times SU(2)$ and we can apply the previous construction based on
this subgroup. In other words, we can deform our theory via
\nref{deform}, which leads to
  the following change in the superpotential\footnote{
There is a  minor subtlety in this case. In order to get the normalization of
 $\beta$ in
 \nref{supkw}, so that its period is $\beta \sim \beta + 1$, we need to define
 the two $U(1)$ to be $n_{1,2} = J^3_{A} \pm J^3_{B} $, where $J^3_{A,B}$ are the
 SU(2) generators that act on the fields $A,B$ and they have a half integer quantization
 condition. The two new combinations we defined have integer quantization when
 they act on gauge invariant operators. These new generators are the ones that
 appear in our general formulas \nref{deform}, etc.}
  \be Tr[A_+B_+A_-B_- - A_-B_+ A_+ B_-] \to Tr[e^{ i
\pi \beta} A_+B_+A_-B_- - e^{- i \pi \beta} A_-B_+ A_+ B_-]
\la{supkw} \ee where $A_\pm$ is in the  ($\bf{N,\bar N}$)
representation and $B_\pm$ is in the ($\bf{\bar N , N}$)
representation of the gauge group $SU(N)\times SU(N)$. This
deformation is exactly marginal \cite{italian}. The reason is that
we have two superpotential parameters, plus the gauge coupling. On
the other hand all beta function equations are proportional to
$\gamma_{A_-} + \gamma_{A_+} + \gamma_{B_-} + \gamma_{B_+} $, so
there is only one constraint. Therefore we have a two dimensional
space of solutions\footnote{ We thank I. Klebanov for this
argument.}. The general analysis of exactly marginal deformations
of this field theory is performed in \cite{italian}. The fact that
there is a corresponding massless field in the gravity solution at
lowest order (in $\beta$) follows from the analysis of the
Kaluza-Klein spectrum in \cite{gubserspec,ferrara}. Another type
of marginal deformation, which preserves a single diagonal $SU(2)$
symmetry was considered in \cite{corrado}.

Let us now turn to
the gravity side, these two $U(1)$ symmetries act geometrically and we can
find the deformed solution by applying the $SL(2,R)$ transformation \nref{sltwo}
to the $AdS_5 \times T^{1,1}$ solution
in \cite{ikew}. This gives\footnote{
We   denote $c_1 \equiv \cos \theta_1 $, $s_2 = \sin \theta_2$, etc.  In order to understand
some of the numerical factors in the following formula it is necessary to remember that we
are applying our general procedure on the two torus parametrized by $
\varphi^{1,2} = {\phi^1 \pm \phi^2 \over 2} $, as we explained in a previous footnote. }
\bea
ds_E^2&=&R^2 G^{-1/4} \left\{ ds^2_{AdS_5} + {s_1^2 s_2^2 \over 324 f} d\psi^2
+\frac{1}{6}(d\theta_1^2+d\theta_2^2) + \right.
\\
& ~& \left. + G\left[ h \left(d \phi_1 + {c_1 c_2 d\phi_2 \over 9h} +
{c_1 d\psi \over 9h} \right)^2
+ {f \over  h} \left( d\phi_2 + { c_2 s_1^2 d\psi \over 54 f} \right)^2
\right]  \right\}
\nonumber\\
e^{2\phi}&=&e^{2 \phi_0} G\nonumber\\
B^{NS}&=&2 \gamma R^4 G f
\left( d \phi_1 + {c_1 c_2 d\phi_2 \over 9h} +
{c_1 d\psi \over 9h}\right)\wedge
\left(d\phi_2 + { c_2 s_1^2 d\psi \over 54 f}   \right)
\nonumber\\
B^{RR}&=&
\gamma \frac{\pi N}{2 }c_1 s_2 d\theta_2 d\psi,
\nonumber\\
F^{(5)}&=&27 \pi N (\omega_{AdS}+*\omega_{AdS})\nonumber\\
&& \la{gdefwk}
G^{-1} \equiv 1+ \gamma^2 4 R^4 f,\qquad ~,~~~~~~~~R^4 = { 27 \over 4}  \pi e^{\phi_0} N
\\
&&h\equiv  \frac{c_1^2}{9}+\frac{s_1^2}{6},\qquad
f\equiv  \frac{1}{54}(c_2^2 s_1^2 + c_1^2 s_2^2 ) +\frac{s_1^2 s_2^2}{36}
\eea
We have presented here the solution when $\beta $ is real ($\beta =\gamma$).
The general solution
can be found in a similar way, see appendix A. In fact,
most of the discussion that we had in the case of $AdS_5\times S^5$ goes through here
with minor modifications. Namely, the discussion on the coulomb branch, as well as some
of the
discussion on the BPS chiral primaries. For example, we can also compute the
Zamolodchikov metric for the marginal deformations to get
\be
S =  { 27 N^2 \over 2^8 \pi^2} \int\left[{\partial \tau_s \partial \bar
\tau_s\over \tau_{2 \, s}^2 } + C
 { | \partial \gamma -
 \tau_s \partial \sigma |^2 \over \tau_{2 \, s} } \right]
\ee
where as in   (\ref{valofc})
\bea
C= \langle g_{0 ,E} \rangle = 4R^4_E\langle f\rangle =4R^4_E\left\langle
\frac{1}{54}(c_2^2 s_1^2 + c_1^2 s_2^2 ) +\frac{s_1^2
s_2^2}{36}\right\rangle=
4R^4_E \frac{5}{243}=\frac{7\pi}{16}N
\eea

Notice that CFT's which are obtained from D-branes at singularities of
toric Calabi-Yau's have three $U(1)$ symmetries. Two of them
 leave the supercharge invariant.
 The gravity solutions
corresponding to these theories \cite{francoetal,MartSparks} have the
form of $AdS_5\times Y_{p,q}$, where the $Y_{p,q}$ spaces were found in \cite{gauntlett}.
 The two
$U(1)$ symmetries are isometries of $Y_{p,q}$. So we can apply  our $SL(2,R)$ method to
deform both the field theory solution and the gravity solution.
These deformations are exactly marginal \cite{italian}. The resulting solutions are
presented in appendix A.

Note that our method based on the transformation \nref{sltwo} is not limited to
conformal field theories. In fact one can also apply it for non-conformal field theories.
The new feature that arises is that the region of validity of the gravity solution
might depend on the radial coordinate, as in \cite{sunny}.
The reason is that through the transformation
\nref{sltwo} a problem arises if the original volume of the two torus becomes too
large in string units. In that case we see that the transformed volume \nref{sltwo}
 becomes very small
so that the gravity solution might become invalid.
This can happen even in a region where the original solution is
non singular. In fact, a simple example is the metric
\nref{rfoursol} which becomes problematic
at large distances.
 In fact, for
the conformal case this was basically the origin of
the first condition in \nref{curv}.
For example, we can consider the gravity solution corresponding to a D-p-brane field
theory for $p<3$ and then apply our method. More specifically,  let us consider the field
theory on D2 branes and  perform this trick with two $U(1)$s in $SU(3)\subset SO(7)$.
We get a supersymmetric theory which is the dimensional reduction of the
$\beta$-deformed theory to 2+1 dimensions.
 The deformation is no longer marginal,
in fact in the UV, which corresponds to weakly coupled Yang Mills, it is a relevant
deformation of dimension 5/2.  The gravity solution
 will involves factors of
\be
G^{-1} = 1 + \gamma^2 f r^4 \sim 1 + \gamma^2{  1 \over r }  ~,~~~~~~~~~~f \sim { 1 \over r^5}
\ee
where $f$ is the harmonic function appearing in the D2 gravity dual \cite{sunny}.
We see that also for strong 't Hooft coupling the deformation becomes important
in the IR. Note that in the supergravity regime this theory is not conformal, nevertheless
we can say that the effects of the deformation become stronger at shorter distance scales,
relative to the effects of the gauge coupling.   It would be nice to see if there
is some valid supergravity description for the IR theory.

Another interesting example is
the cascading theory studied  by Klebanov and Strassler \cite{ikms}.
We can deform both the field theory
and the gravity solution by applying  (\ref{sltwo}),(\ref{deform}). The final
gravity solution looks quite messy and is written in appendix A.
Here let us summarize just a couple of features. Suppose that we make a deformation
with a small but fixed $\gamma$. Then the gravity solution in the UV region becomes
strongly curved if we go far enough. The reason is that in the UV region of the
original solution \cite{ikms} the radius of (the approximately) $T^{1,1}$ is growing,
so that the transformation \nref{sltwo} might generate a highly curved space.
 Another feature of the solution concerns the IR region.
In the IR region the undeformed solution \cite{ikms} has a finite size $S^3$.
Our deformation does not vanish in the IR region of the geometry, it deforms the
three sphere. This might be useful for getting theories that are  closer to
pure ${\cal N}=1$ super-Yang-Mills.

In appendix A we give the general action of the $SL(2,R)$ transformation, so that
it becomes a simple matter to generate new solutions.

Notice that there are some cases where our trick cannot be applied  because some
of the ingredients are not present. For example, if we consider an ${\cal N}=4$
field theory with $SO(N)$ gauge group, then there are no
$\beta$-deformations since the operator in question vanishes identically due to the
anti-symmetry of the matrices in the adjoint representation. Correspondingly, in
the gravity solution there is a two torus, but the $B$ field on this two torus
is projected out so that we cannot form the $\tau$ parameter discussed above.

\section{ Deformations based on $U(1)^3$ symmetries}
\renewcommand{\theequation}{5.\arabic{equation}}
\setcounter{equation}{0}

Suppose that we start with M-theory compactified on a three torus. Then, as we
saw above we have an $SL(3,R)\times SL(2,R)$ symmetry. So far we have been using a
transformation \nref{sltwo} in the $SL(3,R)$ subgroup. A natural question to ask
is whether something interesting can be done with the $SL(2,R)$ subgroup.
In other words, we are interested in applying \nref{sltwo} to the parameter
\be \la{msltwo}
\tau = C_{123} + i \sqrt{G}
\ee
where $C_{123}$ is the value of the $C$ field over the two torus and
$\sqrt{G}$ is the volume of the three torus.

Indeed one can generate new solutions in this fashion.
For example, we can start with the gravity solution describing coincident M2 branes.
We can choose three $U(1)$ symmetries that corresponding to
\bea  \nonumber
(z_1,z_2,z_3,z_4) &\to& (  z_1 ,  z_2,e^{- i \varphi_1 } z_3,e^{i \varphi_1} z_4)
\\ \nonumber
& \to& (z_1,e^{- i \varphi_2} z_2,e^{i \varphi_2} z_3,z_4)
\\ \la{uonetr}
& \to& (e^{i\varphi_3}z_1,e^{- i \varphi_3} z_2, z_3,z_4)
\eea

These symmetries are realized geometrically on $AdS_4\times S^7$.
We see that the deformation based on (\ref{sltwo}) with \nref{msltwo}
gives a new gravity solution
in which the $S^7$ is smoothly deformed. Since the transformations \nref{uonetr}
are embedded in
$SU(4) \subset SO(8)$ so they will preserve two supersymmetries in three dimensions.

On the field theory side this deformation corresponds to turning on an operator in
the fourth spherical harmonic of $SO(8)$ (symmetric traceless representation) which
is also in the fourth fold symmetric representation of $SU(4)$ and is $U(1)^3$ invariant.
There is only one such state, which we can think of as a spherical harmonic of the
form $z_1z_2z_3z_4$. This should be interpreted simply as giving the quantum numbers
of the operator in the field theory. Other marginal deformations were considered in
\cite{kolconf}.
The metric is
\bea
ds_{11}^2&=&G^{-1/3}R^2\left[ { 1 \over 4} ds_{AdS}^2+\sum_{i=1}^4 (d\mu_i^2+G\mu_i^2 d\phi_i^2)
+16\hat \gamma^2 G\mu_1^2\mu_2^2\mu_3^2\mu_4^2(\sum d\phi_i)^2
\right]\nonumber\\
\la{Mthdeform}
F_{(4)}&=&\frac{3}{8}R^3(\omega_{AdS}+16 \hat \gamma s_\theta^5
c_\theta s^2_{2\alpha}s_{2\beta}d\theta d\alpha d\beta d\psi)\\
&&+R^3{\hat\gamma}d\left\{G(\prod \mu_i^2)\left[\frac{d\phi_1 d\phi_2 d\phi_3}{\mu_4^2}+
\frac{d\phi_1 d\phi_3 d\phi_4}{\mu_2^2}-\frac{d\phi_1 d\phi_2 d\phi_4}{\mu_3^2}-
\frac{d\phi_2 d\phi_3 d\phi_4}{\mu_1^2}
\right]\right\}\nonumber\\
\Delta&=&\mu_1^2\mu_2^2\mu_3^2\mu_4^2\sum_1^4 \mu_i^{-2},\qquad
G^{-1}=1+ \hat \gamma^2  \Delta,\quad R=(32\pi^2 N)^{1/6} ~,~~~~~~\hat \gamma \equiv
\gamma R^3
\eea

\section{Conclusions}
\renewcommand{\theequation}{6.\arabic{equation}}
\setcounter{equation}{0}

In this paper we studied a particular deformation of a field theory. This deformation
amounts to a redefinition of the product of fields in the lagrangian by introducing
phases which depend on the order of fields that are charged under a global $U(1)\times U(1)$
symmetry.
Then we can easily do two things:
First, if we know that D-branes on some background give us the undeformed field theory, then
we can deform the background by performing the $SL(2,R)$ transformation \nref{sltwo} in
order to get a new background such that D-branes on this new background give us the
deformed field theory.
Second, if we know the gravity dual of the field theory, and the $U(1)\times U(1)$ symmetry
is realized geometrically, then we can perform the transformation \nref{sltwo} to
get the new background geometry.

This procedure works both for conformal and non-conformal cases. If the original
theory is supersymmetric, then the deformed theory will be supersymmetric if the
$U(1)\times U(1)$ symmetry commutes with the supercharge (i.e. they are not R-symmetries).

It seems that more general dualities might enable us to find very easily other
interesting gravity solutions.

\section*{Acknowledgments }

We would like to thank I. Klebanov and E. Witten for useful discussions.
J.M. would like to thank O. Aharony, D. Berenstein and B. Kol for discussions a
few years ago on this topic. We also want to thank S. Pal for pointing out a few 
typos in the first version of the paper.

This work was supported in part by DOE grant DE-FG02-90ER40542 (JM) and NSF grant PHY-0070928 (OL).

\appendix
\section{Solution generating technique.}
\renewcommand{\theequation}{A.\arabic{equation}}
\setcounter{equation}{0}

In this appendix we summarize the technique which was used to generate supergravity
solutions presented in this paper.

We begin with some solution of 11 dimensional supergravity which
has a $T^3$ symmetry. This KK reduction was considered in
\cite{sezgin,ortin}\footnote{ In \cite{sezgin} they reduced on
$S^3$ and found a gauged supergravity. We can take the limit when
the gauge coupling goes to zero to get the reduction on a torus.
See \cite{ortin} for more general reductions.}
\bea\label{Start11DMetr} ds^2&=& \Delta^{1/3} M_{ab}{\cal
D}\varphi^a {\cal D}\varphi^b
+  \Delta^{-1/6} g_{\mu\nu}dx^\mu dx^\nu\nonumber\\
C^{(3)}&=&\frac{1}{2}(C_{a\mu\nu}{\cal D}\varphi^a\wedge dx^\mu\wedge dx^\nu+
C_{ab\nu}{\cal D}\varphi^a\wedge {\cal D}\varphi^b\wedge dx^\mu)+
\frac{1}{6}C_{\mu\nu\lam}dx^\mu\wedge dx^\nu\wedge dx^\lam
\nonumber\\
&& + C {\cal D} \varphi^1 {\cal D}\varphi^2 {\cal D} \varphi^3
\eea
where the indices $a,b=1,2,3$ and the indices
$\mu, \nu , \cdots$ run over the remaining eight dimensions. The determinant of
the matrix $M$ is one.
Here we use notation ${\cal D}\varphi^a=d\varphi^a+{\cal A}^a_\mu dx^\mu$ for $a=1,2,3$.
Under a general $SL(3,R)$ transformation of coordinates
$(\varphi^1,\varphi^2,\varphi^3)$:
\bea
\left(\begin{array}{c}
\varphi_1\\ \varphi_2\\ \varphi_3
\end{array}\right)'= ( \Lambda^{T} )^{ -1} \left(\begin{array}{c}
\varphi_1\\ \varphi_2\\ \varphi_3
\end{array}\right)
\eea
the fields transform in a following way
\bea
M \rightarrow \Lambda  M\Lambda^T,\qquad&&
\left(\begin{array}{c}
{C}_{23\mu}\\{C}_{31\mu}\\{  C}_{12\mu}
\end{array}\right)\rightarrow
(\Lambda^T)^{-1}
\left(\begin{array}{c}
{C}_{23\mu}\\{C}_{31\mu}\\{  C}_{12\mu}
\end{array}\right),\nonumber\\
\left(\begin{array}{c}
{\cal A}^1_\mu\\{\cal A}^2_\mu\\{\cal A}^3_\mu
\end{array}\right)\rightarrow
(\Lambda^T)^{-1}\left(\begin{array}{c}
{\cal A}^1_\mu\\{\cal A}^2_\mu\\{\cal A}^3_\mu
\end{array}\right),\qquad&&
\left(\begin{array}{c}
{  C}_{1\mu\nu}\\{  C}_{2\mu\nu}\\C_{3\mu\nu}
\end{array}\right)\rightarrow \Lambda
\left(\begin{array}{c}
{  C}_{1\mu\nu}\\{  C}_{2\mu\nu}\\{C}_{3\mu\nu}
\end{array}\right)
\eea In this framework the $SL(3,R)$ symmetry is manifest. Now we
perform a reduction along $\varphi^3$ to produce a solution of
type IIA supergravity. At this step it is convenient to write the matrix
$M$ in \nref{Start11DMetr} as \be M_{ab}{\cal D}\varphi^a {\cal
D}\varphi^b = e^{ - 2 \phi/3} h_{mn} {\cal D}\varphi^m {\cal
D}\varphi^n + e^{ 4 \phi/3} ( {\cal D} \varphi^3 + N_m {\cal D}
\varphi^m)^2 \ee where $m,n = 1,2$ and the determinant of $h_{ab}$
is one. After we find the IIA solution, we   perform a T
duality along $\varphi^1$ to produce a solution of type IIB
supergravity \bea ds_{IIB}^2&=&\frac{1}{ h_{11}} \left[ {1 \over
\sqrt{\Delta} } (D\varphi_1-C D\varphi^2)^2 +\sqrt{\Delta}
(D\varphi^2)^2 \right]
+e^{2\phi/3}g_{\mu\nu}dx^\mu dx^\nu,\nonumber\\
B&=&\frac{h_{12}}{h_{11}}D\varphi^1 \wedge D\varphi^2-
C_{32\mu}D\varphi^2\wedge dx^\mu+D\varphi^1\wedge {\cal A}^1-
\frac{1}{2}C_{3\mu\nu}dx^\mu\wedge dx^\nu
+C_{31\mu}dx^\mu\wedge {\cal A}^1
\nonumber\\
e^{2\Phi}&=&\frac{e^{2\phi}}{h_{11}},\qquad C^{(0)}=N_1\nonumber\\
C^{(2)}&=&-(N_2-\frac{h_{12}}{h_{11}}N_1)D\varphi^1\wedge
D\varphi^2-C_{12\mu}D\varphi^2\wedge dx^\mu - D\varphi^1\wedge
{\cal A}^3_\mu dx^\mu -
\nonumber\\
&&\frac{1}{2} C_{1\mu\nu}dx^\mu\wedge dx^\nu  + C_{31 \mu} dx^\mu
 \wedge {\cal A}^3_\nu
\nonumber\\
C^{(4)}&=&-\left(\frac{1}{2}[C_{2\mu\nu} + 2 C_{32 \mu} {\cal
A}^3_\nu -\frac{h_{12}}{h_{11}} (C_{1\mu\nu} +2  C_{31 \mu} {\cal
A}^3_\nu )]
D\varphi^2 dx^\mu dx^\nu+ \right. \\
&& \left. \frac{1}{6}  (C_{\mu\nu\lam} + 3  C_{3 \mu \nu } {\cal
A}^3_\lambda ) dx^\mu dx^\nu dx^\lam \right)
\wedge  D\varphi^1 \nonumber +
\\ && + d_{\mu_1 \mu_2\mu_3\mu_4} dx^{\mu_1}  dx^{\mu_2} dx^{\mu_3} dx^{\mu_4} +
 \hat d_{\mu_1 \mu_2 \mu_3  }dx^{\mu_1}  dx^{\mu_2} dx^{\mu_3} D\varphi^2  \nonumber \\
&&D\varphi^1=d\varphi^1-C_{31\mu}dx^\mu,\qquad
D\varphi_2=d\varphi^2+{\cal A}^2_\mu dx^\mu
\eea
The forms $d_{\mu_1 \mu_2\mu_3\mu_4}, ~
 \hat d_{\mu_1 \mu_2 \mu_3 } $ are determined by the self duality conditions
 for the five form  field strength. In order to find them, we need to impose the
 equations of motion for $C_{\mu\nu\lambda}$ and $C_{2 \mu \nu}$.
These variables were natural from the eleven dimensional point of view, but they are not
so natural from the point of view of IIB theory. Let us rewrite the solution
as
\bea
ds_{IIB}^2&=&F\left[\frac{1}{\sqrt{\Delta}}
(D\varphi_1-C(D\varphi^2))^2
+\sqrt{\Delta}(D\varphi_2)^2\right]
+ \frac{e^{2\phi/3} }{F^{1/3}}
g_{\mu\nu}dx^\mu dx^\nu,\nonumber\\
B&=&B_{12}(D\varphi^1)\wedge (D\varphi^2)+
\left\{B_{1\mu}(D\varphi^1)+B_{2\mu}(D\varphi^2)\right\}\wedge
dx^\mu
\nonumber\\
&&-\frac{1}{2}A^m_\mu B_{m\nu}dx^\mu\wedge dx^\nu+
\frac{1}{2}{\tilde b}_{\mu\nu}dx^\mu\wedge dx^\nu
\nonumber\\
e^{2\Phi}&=&e^{2\phi},\qquad C^{(0)}=\chi
\nonumber\\
C^{(2)}&=&C_{12}(D\varphi^1)\wedge (D\varphi^2)
+\left\{C_{1\mu}(D\varphi^1)+C_{2\mu}(D\varphi^2)\right\}\wedge
dx^\mu\nonumber\\
&&-\frac{1}{2}A^m_\mu C_{m\nu}dx^\mu\wedge dx^\nu+
\frac{1}{2}{\tilde c}_{\mu\nu}dx^\mu\wedge dx^\nu
\nonumber\\
C^{(4)}&=&-\frac{1}{2}(
{\tilde d}_{\mu\nu}+B_{12}{\tilde c}_{\mu\nu}-
\eps^{mn}B_{m \mu}C_{n\nu}-B_{12}A^m_{\mu}C_{m\nu})
dx^\mu dx^\nu D\varphi^1 D\varphi^2 \nonumber\\
&&+\frac{1}{6}({  C}_{\mu\nu\lam}+3({\tilde b}_{\mu\nu}+
A^1_{\mu}B_{1\nu}-A^2_{\mu}B_{2\nu})C_{1\lam}) dx^\mu dx^\nu
dx^\lam D\varphi^1 + \la{IIBform}
\\ && + d_{\mu_1 \mu_2\mu_3\mu_4} dx^{\mu_1}  dx^{\mu_2} dx^{\mu_3} dx^{\mu_4} +
 \hat d_{\mu_1 \mu_2 \mu_3  } dx^{\mu_1}  dx^{\mu_2} dx^{\mu_3}D\varphi^2
\nonumber \la{cfour} \eea where \bea
D\varphi^2=d\varphi^2+A^2,\qquad D\varphi^1=d\varphi^1+{A}^1 \eea
When we wrote \nref{IIBform} we relabelled various fields and
we have shifted the two forms in
eight dimensions by an $SL(3,R)$ covariant combination of the one forms in eight dimensions.
Then we can deduce the action of $SL(3,R)$ transformations. We
have three objects which transform as vectors: \bea
&&V^{(1)}_\mu=\left(\begin{array}{c} -B_{2\mu}\\A^1_\mu\\C_{2\mu}
\end{array}\right),\quad
V^{(2)}_\mu=\left(\begin{array}{c}
B_{1\mu}\\A^2_\mu\\-C_{1\mu}
\end{array}\right):\quad V^{(i)}_\mu\rightarrow (\Lambda^T)^{-1}V^{(i)}_\mu;
\la{1forms}\\
&&
W_{\mu\nu}=\left(\begin{array}{c}
{\tilde c}_{\mu\nu}\\{\tilde d}_{\mu\nu}\\
{\tilde b}_{\mu\nu}
\end{array}\right)\rightarrow \Lambda W_{\mu\nu}\la{2forms}
\eea
and one matrix
\bea
M=g g^T ~,\quad g^T=
\left(\begin{array}{ccc}
e^{-\phi/3}F^{-1/3}&0&0\\
0&e^{-\phi/3}F^{2/3}&0\\
0&0&e^{2\phi/3}F^{-1/3}
\end{array}\right)
\left(\begin{array}{ccc}
1&B_{12}&0\\
0&1&0\\
\chi&-C_{12}+\chi B_{12}&1
\end{array}\right) \la{demag}
\eea which transforms as \bea M\rightarrow \Lambda  M\Lambda^T
\eea The scalars $\Delta, ~C$ as well as the three form ${
C}_{\mu\nu\lam}$ stay invariant under these $SL(3,R)$
transformations. Under $SL(2,R)$ transformations  $ C + i
\sqrt{\Delta}$ transform as a $\tau $ parameter. Then the one
forms \nref{1forms} transform into each other, the two forms
\nref{2forms} remain invariant and the fourform field strength in
eight dimensions coming from $ C_{\mu \nu \delta}$ in the
last line of \nref{cfour} transforms into its magnetic dual in
eight dimensions. The metric $g_{\mu \nu}$ is the Einstein metric in eight
dimensions and does not change under any of the transformations.

Let us consider a particular example of the $SL(3,R)$
 transformation. We begin with
geometry which has only metric and ${\tilde d}_{\mu\nu}$ excited, with all other
fields set to zero, including the dilaton. This
implies that the five--form field strength can be written as
\bea
F_{(5)}={\tilde F}_{(5)}+~^*{\tilde F}_{(5)}
\eea
where ${\tilde F}_{(5)}$ has no indices along the torus $(\varphi^1,\varphi^2)$.
Let us apply the transformation with\footnote{
The matrix $h_3$ used in \nref{slthree} (with $h_2 = 1$) is equivalent to \nref{lammat}
after interchanging the first two rows and columns. }
\bea \la{lammat}
\Lambda=\left(\begin{array}{ccc}
1& \gamma&0\\
 0&1& 0 \\
0&\sigma &1
\end{array}\right)
\eea
Then we find the transformed fields
\bea
{g'}^T&=&\left(\begin{array}{ccc}
\frac{F^{-1/3}}{\sqrt{GH}}&0&0\\
0&F^{2/3}\sqrt{G}&0\\
0&0&F^{-1/3}\sqrt{H}
\end{array}\right)
\left(\begin{array}{ccc}
1&\gamma F^2 G &0\\
0&1&0\\
\gamma \sigma \frac{F^2}{H} & \sigma {F^2 \over H} &1
\end{array}\right)\nonumber
\eea
By comparing this with \nref{demag} we can find the new fields
\bea
B'_{12}&=&\gamma F^2 G,\quad
C'_{12}=-\sigma F^2 G ,\quad
\chi'=\gamma \sigma \frac{F^2}{H}
\quad F' = F G \sqrt{H}, \quad e^{2 \phi'} = G H^2 \nonumber\\
&&G^{-1}\equiv 1+(\gamma^2+\sigma^2)F^2,\quad
H=1+\sigma^2 F^2
\eea
and the new geometry
\bea
ds_{IIB}^2&=&FGH^{1/2}
\left[\frac{1}{\sqrt{\Delta}}
(D\varphi_1-C(D\varphi^2))^2
+\sqrt{\Delta}(D\varphi_2)^2\right]
+H^{1/2}F^{-1/3}
g_{\mu\nu}dx^\mu dx^\nu,\nonumber\\
B&=&\gamma F^2 G(D\varphi^1)\wedge (D\varphi^2)+
\frac{\sigma}{2}{\tilde d}_{\mu\nu}dx^\mu\wedge dx^\nu,
\qquad
e^{2\phi}=H^2 G
\nonumber\\
C^{(2)}&=&-\sigma F^2 G(D\varphi^1)\wedge (D\varphi^2)+
\frac{\gamma}{2}{\tilde d}_{\mu\nu}dx^\mu\wedge dx^\nu,\qquad
\chi=\gamma\sigma F^2 H^{-1},\nonumber\\
F^{(5)}&=&{\tilde F}_{(5)}+~^*{\tilde F}_{(5)} \la{gennew}
\eea
Notice that ${\tilde F}_{(5)}$ is the same as before, but the star is
now taken with the new metric.

As an example we consider an application this of
procedure to flat space. We begin with metric on
$R^{10}$ which we write in a form
\bea
ds^2&=&\eta_{\mu\nu}dx^\mu dx^\nu+\sum_{i=1}^3 dr_i^2+r_1^2 (d\psi -d\varphi_2)^2+
r_2^2(d\psi+ d\varphi_1+d\varphi_2)^2+ r_3^2 (d\psi -d\varphi_1)^2\nonumber\\
&=&\eta_{\mu\nu}dx^\mu dx^\nu+\sum dr_i^2+
(r_2^2+r_3^2)\left(D\varphi_1+\frac{r_2^2}{r_2^2+r_3^2}D\varphi_2\right)^2+
\frac{g_0}{r_2^2+r_3^2}(D\varphi_2)^2\nonumber\\
&&+\frac{9 r_1^2 r_2^2 r_3^2}{g_0}d\psi^2\nonumber
\eea
where we defined
\bea
&&g_0\equiv r_1^2r_2^2+r_1^2r_3^2+r_2^2r_3^2,\nonumber\\
&&D\varphi_1=d\varphi_1-d\psi+\frac{3r_1^2r_2^2 }{  g_0}d\psi,\qquad
D\varphi_2=d\varphi_1-d\psi+\frac{3 r_2^2r_3^2}{  g_0}d\psi,
\eea

Then transformation with parameter $\gamma$ gives a new solution of type IIB supergravity
\bea
ds_{IIB}^2&=&\eta_{\mu\nu}dx^\mu dx^\nu+\sum dr_i^2+
\nonumber\\
&&G\left[  r_1^2 (d\psi -d\varphi_1)^2+
r_2^2(d\psi+ d\varphi_1+d\varphi_2)^2+ r_3^2 (d\psi -d\varphi_2)^2   +
9\gamma^2 g_0  r_1^2 r_2^2 r_3^2 d\psi^2 \right] \nonumber\\
B&=&\gamma g_0 G (D\varphi^1)\wedge (D\varphi^2)
\qquad
e^{2\Phi}=G,\qquad G^{-1}\equiv 1+\gamma^2g_0
\nonumber
\eea
This background preserves two supersymmetries in four dimensions, namely  the
two that transform as a singlet of $SU(3)$ acting on the three complex coordinates
of $R^6$.
If we place D3 branes at the origin of this space, $r_i=0$, and take a low energy
limit we find the $\beta$-deformation of ${\cal N}=4$ super Yang Mills.
Application of this procedure to
$AdS_5\times S^5$ gives the solution (\ref{metrgen}).

Let us discus now the regularity of the transformed solution.
Suppose that we start with a metric that is non-singular as a ten dimensional
theory. When do we get a non-singular solution?.
In principle there are a variety of things that could go wrong.
For example, the original theory could be  such that $\tau$, in \nref{taudef}
 is only defined up to
gauge transformations  of the $B$ field. Then the final geometry
will not be well defined. So $\tau$ has to be globally well defined and
should be such that $\tau_{1} \to 0$ when $\tau_2 \to 0 $.
In the case that the original $B$ field is non-zero we also should worry
about the components of the $B$ field that are vectors in
eight dimensions. In eight dimensions these vectors can have Wilson lines, which
if integer, are allowed by the regularity conditions. However, under the
transformation the field $A^1_\mu \to A^1_\mu + \gamma B_{2 \mu}$ and this can lead
to non-integer Wilson lines. So this is another thing that needs to be checked.
When we check these properties we are allowed to redefine the coordinates
$\varphi^m$ in such a way to make our job easier.
This amounts to gauge transformations of the $A^m_\mu$ fields. It turns out that performing
this transformations before or after the $SL(2,R)$ transformation does not change
the final answer. So the conclusion is that $A$ need not be globally defined and it
need not have vanishing Wilson lines, as long as the original metric is regular and the
$B_\mu$ and $C_\mu$ gauge fields are globally well defined one forms.
These remarks are useful for checking the regularity of the deformed Klebanov-Strassler
solution.
Another potential problem is the fact that the last terms in the $B$ and $C^{(2)}$ fields in
\nref{gennew} might lead to fluxes that are not properly quantized. An example were
we would run into a problem arises if we have a $T^2 \times S^3$ with $N$ units
of flux of $F_5$ in the original geometry. Then, for general $\gamma$,
 we would have a non-quantized
 $dC^{(2)}$ flux on the $S^3$, which is not allowed. We have checked on a case by
 case basis that this does not happen for our solutions.

\subsection{Gravity dual of the conifold conformal field theory.}

It is straightforward to apply the procedure outlined above to $AdS_5\times T^{1,1}$.
\bea
{ds_E \over R^2_E}^2&=&ds_{AdS}^2+\frac{1}{9}(d\psi+c_{1} d\phi_1+c_{2} d\phi_2)^2+
\frac{1}{6}(s^2_1 d\phi_1^2+s^2_2 d\phi_2^2+d\theta_1^2+d\theta_2^2)\nonumber\\
F^{(5)}&=& 4 R^4_E (\omega_{AdS}+^*\omega_{AdS})
\eea
We rewrite it to make the $T^2$ part more explicit
\bea
{ ds^2 \over R^2} &=& h \left(d \phi_1 + {c_1 c_2 d\phi_2 \over 9h} +
{c_1 d\psi \over 9h} \right)^2
+ {f \over h} \left( d\phi_2 + { c_2 s_1^2 d\psi \over 54f} \right)^2
+ {s_1^2 s_2^2 \over 324 f} d\psi^2\nonumber\\
&&+\frac{1}{6}(d\theta_1^2+d\theta_2^2)+ds^2_{AdS}\\
&&h\equiv  \frac{c_1^2}{9}+\frac{s_1^2}{6},\qquad
f\equiv  \frac{1}{54}(c_2^2 s_1^2 + c_1^2 s_2^2 ) +\frac{s_1^2 s_2^2}{36}
\eea
This implies that the deformed metric is simply
\bea
{ds_E^2 \over R_E^2}&=&G^{3/4}\left[ h \left(d \phi_1 + {c_1 c_2 d\phi_2 \over 9h} +
{c_1 d\psi \over 9h} \right)^2
+ {f \over h} \left( d\phi_2 + { c_2 s_1^2 d\psi \over 54f} \right)^2
\right]\\
&&+G^{-1/4}\left[{s_1^2 s_2^2 \over 324 f} d\psi^2
+\frac{1}{6}(d\theta_1^2+d\theta_2^2)+ds^2_{AdS}\right]
\nonumber\\
{B \over R_E^2}&=& \hat \gamma G f
\left(d\phi_1 +\frac{c_1c_2 d\phi_2 }{ 9 h} + { c_1 d\psi \over 9 h} \right)\wedge
\left(d\phi_2+{c_2 s_1^2 \over 54 f}d\psi \right)+
\frac{\hat \sigma}{27}c_1 s_2 d\theta_2 d\psi,
\nonumber\\
{C^{(2)} \over R_E^2}&=&- \hat \sigma G f
\left( d\phi_1 +\frac{c_1c_2 d\phi_2 }{ 9 h} + { c_1 d\psi \over 9 h}\right)\wedge
\left(d\phi_2+{c_2 s_1^2 \over 54f}d\psi \right)+
\frac{\hat \gamma}{27}c_1 s_2 d\theta_2 d\psi,
\nonumber\\
{ F^{(5)} \over R_E^4} &=& 4 (\omega_{AdS}+*\omega_{AdS})\nonumber\\
  \quad
e^{2\Phi}&=&G H^2 ~,~~~~~~~~\quad \chi= \hat \gamma \hat \sigma  f H^{-1}
\\
G^{-1}&\equiv & 1+(\hat \sigma^2+\hat \gamma^2)f,\qquad
H \equiv 1+\hat \sigma^2 f ~,~~~~~\hat \gamma \equiv 2 \gamma R_E^2 ~,~~~~ \hat \sigma
 \equiv 2 \sigma R_E^2
\eea
Here we presented the solution when $\tau_{s} = i $. For general $\tau_{s}$ we
get a solution similar to \nref{metrgen}, so that it is hopefully obvious to
the reader how to introduce the $\tau_s$ dependence.

\subsection{ Marginal deformations of theories associated to the Y$^{p,q}$ manifolds}

As another example we consider the deformations of recently discovered Sasaki--Einstein
spaces Y$^{p,q}$ \cite{gauntlett,MartSparks}
\bea
ds^2&=&ds_{AdS_5}+\frac{1-y}{6}(d\theta^2+s_\theta^2 d\phi^2)+\frac{1}{w(y)q(y)}
dy^2+\frac{q(y)}{9}(d\psi-c_\theta d\phi)^2 \nonumber\\
&&+  {w(y)}\left[\ell d\alpha +f(y) (d\psi-c_\theta d\phi)\right]^2
\eea
where
\bea
&&w(y)=\frac{2(a-y^2)}{1-y},\qquad
 q(y)=\frac{a-3y^2+2y^3}{a-y^2}, \qquad
f(y)=\frac{a-2y+y^2}{6(a-y^2)}\\
&&\ell=\frac{q}{3q^2-2p^2+p(4p^2-3q^2)^{1/2}} \eea where we are
using the same notation is in \cite{MartSparks}, except that our
$\alpha$ has period $2\pi$ (and we have set their $c=1$).
 We
identify the two $U(1)$ symmetries as the symmetries shifting
$\alpha$ and shifting $\phi$. Indeed the holomorphic three form in
\cite{MartSparks} is invariant under such shifts. Rearranging the
metric to make the $T^2$ more explicit, we get \bea
ds^2&=&ds_{AdS_5}+\frac{1-y}{6} d\theta^2+\frac{1}{w(y) q(y)}
dy^2+\frac{ q(y)(1-y)s_\theta^2 d\psi^2}{3(2 q(y)c_\theta^2+3s_\theta^2(1-y))}\\
&&+w(y)\left[\ell d\alpha+f(y)(d\psi-c_\theta d\phi)\right]^2
+\frac{2q(y) c_\theta^2+3(1-y)s_\theta^2}{18}
\left(d\phi-\frac{2q(y)c_\theta d\psi}{2q(y)c_\theta^2+3s_\theta^2(1-y)}\right)^2 \nonumber\\
F_{(5)}&= &\omega_{AdS}+\frac{\ell}{18}(1-y)s_\theta~ d\theta dy d\psi d\alpha d\phi
\eea
Then application of our procedure gives the deformed solution\footnote{Here we restored the
scales associated with the geometry.}
\bea
ds_E^2&=&R_E^2 G^{-1/4}\left\{\left[ds_{AdS_5}+\frac{1-y}{6} d\theta^2+\frac{1}{w(y)q(y)}
dy^2+
\frac{q(y)\ell^2}{27}G(\gamma^2+\sigma^2)s_\theta^2(a-y^2) d\psi^2\right]\right. \nonumber\\
&&+G\left.\left[\frac{1-y}{6} s_\theta^2 d\phi^2+
\frac{q(y)}{9}(d\psi-c_\theta d\phi)^2 +
{w(y)}\left[\ell d\alpha +f(y) (d\psi-c_\theta d\phi)\right]^2\right]\right\}\\
B&=&\gamma g_0 R_E^4 G(D\alpha)\wedge (D\phi)+{ \pi N \over {\cal V}}
\frac{8\sigma\ell}{9}(1-y)
c_\theta~ dy d\psi,
\qquad
e^{2\phi}=H^2 G
\nonumber\\
C^{(2)}&=&
-\sigma g_0 R_E^4 G(D\alpha)\wedge (D\phi)+{ \pi N \over {\cal V}}
\frac{8\gamma\ell}{9}(1-y)
c_\theta~ dy d\psi,\qquad
\chi= \gamma\sigma g_0 H^{-1},\nonumber\\
F^{(5)}&=&{16\pi N \over {\cal V}}(\omega_{AdS}+^*{\omega}_{AdS})\\
&&G^{-1}=1+ (\gamma^2+\sigma^2)g_0,\quad H=1+ \sigma^2 g_0, \nonumber \\
&&
g_0=\frac{2q (y)c_\theta^2+3(1-y)s_\theta^2}{9(1-y)}(a-y^2)\ell^2\nonumber\\
&&D\alpha=d\alpha+\frac{1}{\ell}\frac{3f(y)s^2_\theta (1-y)d\psi}{2q(y)c_\theta^2+3s_\theta^2(1-y)},\qquad
D\phi=d\phi-\frac{2q(y)c_\theta d\psi}{2q(y)c_\theta^2+3s_\theta^2(1-y)}
\eea
Here we defined the ratio of volumes ${\cal V}$ (see \cite{gauntlett} for
details)
\bea
&&{\cal V}\equiv \frac{\mbox{vol} (Y^{p,q})}{\mbox{vol} (S^5)} =
     \frac{q^2[2p+(4p^2-3q^2)^{1/2}]}{3p^2[3q^2-2p^2+p(4p^2-3q^2)^{1/2}]},\qquad
R_E^4={ 4\pi N \over {\cal V}},\nonumber
\eea

\subsection{Deformations of the Klebanov-Strassler solution}

Our procedure can be applied to non--conformal theories as well. An example of the
gravity solution corresponding to such theory is the Klebanov--Strassler background
\cite{ikms}\footnote{We use the same notation as \cite{ikms}, reader should consult that paper
for the explicit form of $f,k,h,K$. Notice however that to have the same units of flux as we use
in the rest of the paper, we redefined a parameter $M$ compared to \cite{ikms}:
$M_{our}=\frac{1}{2}M_{KS}$. Our normalization is also the one
 used in \cite{conifoldreview}.}
\bea
ds^2&=&h^{-1/2}m^2 dx_m dx_m+h^{1/2}\frac{3^{1/3}}{2^{4/3}}K\left[
\frac{1}{3K^3}(d\tau^2+(g_5)^2)+\cosh^2\frac{\tau}{2}[(g_3)^2+(g_4)^2]
\right.\nonumber\\
&&\left.+
\sinh^2\frac{\tau}{2}[(g_1)^2+(g_2)^2]
\right]\\
G_{(3)}&=&\frac{M}{2}\left[g_5\wedge g_3\wedge g_4+d\{
F(g_1\wedge g_3+g_2\wedge g_4)\}\right]\nonumber\\
B&=&\frac{g_s M}{2}\left\{f g_1\wedge g_2+k g_3\wedge g_4\right]
\nonumber
\eea
where the one--forms $g_i$ are defined by
\bea
&&
g_1=\frac{1}{\sqrt{2}}(-s_1 d\phi_1-c_\psi s_2 d\phi_2+s_\psi d\theta_2),
\qquad
g_2=\frac{1}{\sqrt{2}}(d\theta_1-s_\psi s_2 d\phi_2-c_\psi d\theta_2),
\nonumber\\
&&
g_3=\frac{1}{\sqrt{2}}(-s_1 d\phi_1+c_\psi s_2 d\phi_2-s_\psi d\theta_2),
\qquad
g_4=\frac{1}{\sqrt{2}}(d\theta_1+s_\psi s_2 d\phi_2+c_\psi d\theta_2),
\nonumber\\
&&g_5=d\psi+c_1 d\phi_1+c_2 d\phi_2
\eea
We can now use the procedure outlined above to construct the deformed
solution. For example, performing $SL(3,R)$ transformation with parameter
$\gamma$, we find ingredients of the modified metric
\bea
B_{12}&=&\frac{g_s M}{4}(f-k)s_1 s_2 s_\psi,\\
\eea
Then we find
\bea
&&(A^1_\mu)'=(1+\gamma B_{12})A_\mu^1+
\frac{\gamma g_s M}{4}s_2 [(f+k)d\theta_2+c_\psi(k-f)d\theta_1]\nonumber\\
&&(A^2_\mu)'=(1+\gamma B_{12})A_\mu^2+
\frac{\gamma g_s M}{4}s_1[(f+k)d\theta_1+c_\psi(k-f)d\theta_2]\nonumber\\
&&G^{-1}=(1+\gamma B_{12})^2+\gamma^2 \mbox{det}~g
\eea
Here $\sqrt{\mbox{det}~g}$ is a volume of the 2--torus, and we will not
give the explicit forms for $A^1, ~A^2$. The important point we want to
convey is that the original $B$ field will appear in the deformed metric.
The transformed metric can be written in a form
\bea
ds^2&=&h^{-1/2}m^2 dx_m dx_m+\frac{h^{1/2} d\tau^2}{3^{2/3}2^{4/3}K^2} +
 \\ &&+h^{1/2}\frac{3^{1/3}}{2^{4/3}}KG\left[
\frac{(g_5)^2}{3K^3}+\cosh^2\frac{\tau}{2}[(g_3)^2+(g_4)^2]
 + \sinh^2\frac{\tau}{2}[(g_1)^2+(g_2)^2]
\right] + \nonumber\\
&&+h^{1/2}(1-G)\frac{3^{1/3}}{2^{4/3}}K\left\{
\frac{s_1^2 s_2^2\sinh^2\tau}{2 H}
(\cosh\tau (d\theta_1^2+d\theta_2^2)+2c_\psi d\theta_1 d\theta_2)\right. +
\nonumber\\
&&+\frac{H}{12 K^3{\cal F}} \times  \nonumber \\
&&\left.\times[d\psi-
\frac{ s_\psi s_1 s_2}{H} (c_1 c_\psi s_2+c_2 s_1\cosh\tau)d\theta_1-
\frac{ s_\psi s_1 s_2}{H} (c_2 c_\psi s_1+c_1 s_2\cosh\tau)d\theta_2
] ^2\right\}\nonumber
\eea
Here we introduced various functions
\bea
&&H \equiv s_1^2 s_2^2(\cosh^2 \tau -c_\psi^2)\nonumber\\
&& {\cal F} \equiv
 { 1 \over 6 K^3} \left[ (c_2^2 s_1^2+c_1^2 s_2^2)\cosh\tau+ 2 s_1 s_2 c_1c_2c_\psi \right]+ { 1 \over 4}
 s^2_1s^2_2 (\cosh^2 \tau - \cos^2 \psi),\nonumber\\
&&G^{-1} \equiv \left(1+
\frac{\gamma g_s M}{4}(f-k)s_1 s_2 s_\psi\right)^2+
\gamma^2 h\frac{3^{2/3}}{2^{8/3}}K^2{\cal F}
\nonumber
\eea
We first look at large values of $\tau$. At  leading order we get
\bea
G^{-1}&\sim&1+\gamma^2\left(\frac{3}{2}\right)^{2/3}\frac{\alpha\tau}{64}
(c_2^2s_1^2+c_1^2 s_2^2+\frac{3}{2}s_1^2 s_2^2)
\eea
and the metric becomes
\bea
ds^2&=&h^{-1/2}m^2 dx_m dx_m+\frac{h^{1/2} d\tau^2}{3^{2/3}2^{4/3}K^2}\nonumber\\
&&+h^{1/2}e^{\tau}\frac{3^{1/3}}{2^{4/3}}KG
\left[
\frac{1}{6}(g_5)^2+\frac{1}{4}[(g_1)^2+(g_2)^2+(g_3)^2+(g_4)^2]
\right.\nonumber\\
&&\qquad \left.
+ \gamma^2\left(\frac{3}{2}\right)^{2/3}\frac{\alpha\tau}{256}
\left\{
(c_2^2s_1^2+c_1^2 s_2^2+\frac{3}{2}s_1^2 s_2^2)(d\theta_1^2+d\theta_2^2)+
s_1^2 s_2^2 d\psi^2\right\}\right]\nonumber
\eea
This metric becomes highly curved for large $\tau$.
Notice that in contrast to the conformal case, the $\gamma$--deformation
grows with $\tau$. Note also that $B_{12} $ decreases for large $\tau$.

At $\tau=0$ we get the approximate expressions
\bea
&&K=(2/3)^{1/3},\quad H=s_1^2s_2^2 s_\psi^2,\nonumber\\
&&{\cal F}=\frac{1}{4}(c_2^2s_1^2+c_1^2s_2^2+2s_1s_2c_1c_2c_\psi+
s_1^2 s_2^2 s_\psi^2)
\quad G^{-1}=1+\frac{\gamma^2 h}{4}{\cal F} \la{gconif}
\eea
and the metric becomes
\bea
ds^2&=&h^{-1/2}m^2 dx_m dx_m+\frac{h^{1/2} d\tau^2}{3^{2/3}2^{4/3}K^2}+{ h^{1/2}G \over 2}
\left[
\frac{1}{2}(g_5)^2+(g_3)^2+(g_4)^2 + \right.
\nonumber\\
&&\left.+\frac{\gamma^2 h}{32}
(\frac{1}{\sqrt{2}}g_5+c_1 s_2 s_\psi g_3-
(c_2 s_1+c_1 s_2 c_\psi)g_4)^2\right] \la{ultima}
\eea
We see that sphere $S^3$ which is located at the origin of $\tau$ is deformed.
It can be checked that $G$ in \nref{gconif} is a function of only the 3-sphere
coordinates. Similarly one can check that the last term in \nref{ultima} depends only
on the angles on $S^3$. In other words the metric on the three sphere is deformed as
it would be by doing the $SL(2,R)$ transformation on two commuting isometries. We can
define new coordinates $\tilde \phi_1 , \tilde \theta , \tilde \phi_2$ through \cite{minasian}
\bea
e^{ i \sigma^3 {\tilde \phi_1 \over 2}} e^{ i \sigma^2 {\tilde \theta \over 2}}
e^{ i \sigma^3 { \tilde \phi_2 \over 2}} =
e^{ i \sigma^3 {  \phi_1 \over 2}} e^{ i \sigma^2 {  \theta_1 \over 2}}
e^{ i \sigma^3 {  \psi \over 2}}  e^{ i \sigma^2 {\tilde \theta_2 \over 2}}
e^{ i \sigma^3 {   \phi_2 \over 2}}
\eea
Note that under shifts of $\phi_i$, the
$\tilde \phi_i$ shift in the same way.
Then we can write the metric \nref{ultima} as
\bea
ds^2&=&h^{-1/2}m^2 dx_m dx_m+\frac{h^{1/2} d\tau^2}{3^{2/3}2^{4/3}K^2}+
\\ && + { h^{1/2} \over 4}\left\{
G\left[ (d\tilde \phi_1 + \cos\tilde \theta d\tilde \phi_2)^2 +
\sin^2\tilde \theta d\tilde \phi_2^2   \right] + d\tilde \theta^2
\right\}
\\
&& G^{-1} = 1 + \gamma^2{ h \over 4} \sin^2\tilde \theta
\eea
Notice that only the $\gamma$ transformation leads to a non-singular metric. If we tried
to do a $\sigma$ transformation on this three sphere we would run into trouble since
there is $H_{RR}$ flux on it and $C^{(2)}_{12}$ would not be well defined.

\subsection{Marginal deformation of $AdS_4\times S^7$.}

So far we have generated several solutions of type IIB supergravity using
the lift
to eleven dimensions and $SL(3,R)$ group of eleven dimensional
supergravity on the
three--torus. One can also produce new solutions of eleven dimensional
supergravity by
reducing to type IIB and using $SL(2,R)$ symmetry there. To illustrate
this procedure we
consider the example of $AdS_4\times S^7$
\bea
ds^2&=&{ 1 \over 4} ds_{AdS}^2+d\Omega_7^2,\qquad
F_{(4)}=\frac{3}{2^3}\omega_{AdS_4}
\eea
Let us parameterize the sphere as
\bea
d\Omega_7^2&=&d\theta^2+s^2_\theta(d\alpha^2+s_\alpha^2 d\beta^2)
+c^2_\theta d\phi_1^2
+s^2_\theta\left[c^2_\alpha d\phi_2^2+ s^2_\alpha (c^2_\beta
d\phi_3^2+s^2_\beta d\phi_4^2)
\right]
\eea
and introduce new angles
\bea
\phi_1=\psi+\varphi_3,\qquad
\phi_2=\psi-\varphi_3-\varphi_2,\qquad
\phi_3=\psi+\varphi_2-\varphi_1,\qquad
\phi_4=\psi+\varphi_1,
\eea
Then we can rewrite the geometry in a form (\ref{Start11DMetr})
with following ingredients:
\bea
&&\Delta^{1/3}e^{4\phi/3}=c_\theta^2+s^2_\theta c^2_{\alpha},\quad
\Delta=s^4_\theta s^2_\alpha
(c_\theta^2 c^2_\alpha+s^2_{\alpha}s^2_{\beta}c^2_{\beta}
(c^2_\theta+s^2_\theta c^2_\alpha))
=\mu_1^2\mu_2^2\mu_3^2\mu_4^2\sum_1^4 \mu_i^{-2},
\nonumber\\
&&\Delta^{1/3}h_{mn}D\varphi^m D\varphi^n=e^{2\phi/3}\left[
s^2_\theta s^2_\alpha(D\varphi_1-c^2_{\beta} D\varphi_2)^2+
s^2_\theta\frac{c_\theta^2 c^2_\alpha+s^2_{\alpha}s^2_{\beta}c_\beta^2
(c^2_\theta+s^2_\theta c^2_\alpha)}{c^2_\theta+s^2_\theta c^2_\alpha}
D\varphi_2^2\right]\nonumber\\
&&
{\cal A}^1=
\frac{-4(1+2c_{2\beta})c_\theta^2 c^2_\alpha+s^2_{\alpha}s^2_{2\beta}
(c^2_\theta+s^2_\theta c^2_\alpha)}{4c_\theta^2 c^2_\alpha+
s^2_{\alpha}s^2_{2\beta}
(c^2_\theta+s^2_\theta c^2_\alpha)}d\psi,\qquad
{\cal A}^2=2\frac{-4c_\theta^2 c^2_\alpha+s^2_{\alpha}s^2_{2\beta}
(c^2_\theta+s^2_\theta c^2_\alpha)}{4c_\theta^2 c^2_\alpha+
s^2_{\alpha}s^2_{2\beta}
(c^2_\theta+s^2_\theta c^2_\alpha)}d\psi,\qquad
\nonumber
\eea
\bea
&&N_1=0,\quad
N_2=\frac{s^2_\theta c^2_\alpha}{c^2_\theta+s^2_\theta c^2_\alpha},
\quad {\cal A}^3=\left(1-
\frac{4 s^2_{\alpha}s^2_{2\beta}s^2_\theta c^2_{\alpha}}{
4c_\theta^2 c^2_\alpha+
s^2_{\alpha}s^2_{2\beta}
(c^2_\theta+s^2_\theta c^2_\alpha)}\right)d\psi\nonumber\\
&&\Delta^{-1/6}g_{\mu\nu}dx^\mu dx^\nu=
{ 1 \over 4} ds_{AdS}^2+d\theta^2+s^2_\theta(d\alpha^2+s_\alpha^2 d\beta^2)
+\frac{s^2_{2\theta} s^2_{2\alpha}s^2_{2\beta}d\psi^2}{
4c_\theta^2 c^2_\alpha+
s^2_{\alpha}s^2_{2\beta}
(c^2_\theta+s^2_\theta c^2_\alpha)}\nonumber
\eea
The geometry of Type IIB becomes:
\bea
ds_{IIB}^2&=&\frac{1}{h_{11}}\frac{(d\varphi_1)^2}{\sqrt{\Delta}}
+\frac{\sqrt{\Delta}}{h_{11}}
(D\varphi^2)^2
+e^{2\phi/3}g_{\mu\nu}dx^\mu dx^\nu,\nonumber\\
B&=&-c^2_{\beta}d\varphi^1 \wedge D\varphi^2+d\varphi^1\wedge {\cal A}^1
\nonumber\\
e^{2\Phi}&=&\frac{e^{2\phi}}{h_{11}},\qquad C^{(0)}=0\nonumber\\
C^{(2)}&=&-\frac{ s^2_\theta c^2_\alpha}{
c^2_\theta+s^2_\theta c^2_\alpha}
d\varphi^1\wedge D\varphi^2-
d\varphi^1\wedge {\cal A}^3, \\
C^{(4)}&=&- \frac{3}{8}
\left(  w_3d\varphi^1 + \sqrt{\Delta} {\hat w}_3 d\varphi^2 \right)
~,\qquad dw_3 = \omega_{AdS_4},\quad \nonumber
d{\hat w}_3=\sqrt{\Delta}~^*_8 dw_3
\eea
Making $SL(2,R)$ transformation in the $\varphi^1$-$\varphi^2$ plane:
\bea
\left(\begin{array}{c}
\varphi^1\\ \varphi^2
\end{array}\right)\rightarrow
\left(\begin{array}{cc}
{\hat\alpha}&{\hat\beta}\\ {\hat\gamma}&{\hat\delta}
\end{array}\right)
\left(\begin{array}{c}
\varphi^1\\ \varphi^2
\end{array}\right)
\eea
we find a more general geometry and we can read off the
modified ingredients of eleven dimensional metric. First we look at
\bea
&&\Delta'=\frac{\Delta}{({\hat\alpha}^2+{\hat\gamma}^2\Delta)^2},\qquad
h_{11}'=h_{11},\quad
C'_{123}=-\frac{{\hat\alpha}{\hat\beta}+{\hat\gamma}{\hat\delta}\Delta}{
{\hat\alpha}^2+{\hat\gamma}^2\Delta}
\eea
The metric should be regular as $\Delta\rightarrow 0$, this selects the
value ${\hat\alpha}=1$. Also at $\Delta=0$ the 2--torus contracts, so the components
of tensor fields along this torus should go to zero. In particular,
$C'_{123}$ should go to zero at these points which leads to ${\hat\beta}=0$.
Setting ${\hat\alpha}={\hat\delta}=1$, ${\hat\beta}=0$, we get the expressions for the transformed
quantities:
\bea
&&\Delta'=\frac{\Delta}{(1+{\hat\gamma}^2\Delta)^2},\quad
h_{11}'=h_{11},\quad
e^{2\phi'}=e^{2\phi},\quad
g'_{\mu\nu}=g_{\mu\nu},\nonumber\\
&&({\cal A}^a)'={\cal A}^a,\quad N_m'=N_m,\quad
C'_{123}=-\frac{{\hat\gamma}\Delta}{1+{\hat\gamma}^2\Delta},\quad
C'_{12\mu}=0,\nonumber\\
&&C'_{1\mu\nu}=C'_{2\mu\nu}=0,\qquad
C'_{1\mu}=0,\quad C'_{\mu\nu}=0.
\eea
Substituting this into the eleven dimensional geometry, we get
\bea
ds_{11}^2&=&G^{-1/3}\left[ { 1 \over 4} ds_{AdS}^2+\sum (d\mu_i^2+G\mu_i^2 d\phi_i^2)
+16{\hat\gamma}^2 G\mu_1^2\mu_2^2\mu_3^2\mu_4^2(\sum d\phi_i)^2
\right]\nonumber\\
F_{(4)}&=&\frac{3}{8}(\omega_{AdS}+16{\hat\gamma} s_\theta^5
c_\theta s^2_{2\alpha}s_{2\beta}d\theta d\alpha d\beta d\psi)
-{\hat\gamma}d\left\{\Delta G D\varphi_1 D\varphi_2 D\varphi_3\right\}\\
\Delta&=&\mu_1^2\mu_2^2\mu_3^2\mu_4^2\sum_1^4 \mu_i^{-2},\qquad
G^{-1}=1+{\hat\gamma}^2\Delta
\eea
Notice that
\bea
&&D\varphi_1=d\varphi_1+d\psi-\frac{s^4_\theta s^2_\alpha}{\Delta}4c_\beta^2c_\theta^2
c_\alpha^2 d\psi=d\phi_4-\frac{4\prod \mu_i^2}{\Delta}
\frac{d\psi}{\mu_4^2},\\
&&D\varphi_2=d\varphi_2+2d\psi-\frac{4}{\Delta}s^4_\theta s^2_\alpha c_\theta^2 c_\alpha^2
d\psi=
d\phi_3+d\phi_4-
\frac{4\prod \mu_i^2}{\Delta}\left(\frac{d\psi}{\mu_3^2}+\frac{d\psi}{\mu_4^2}\right),
\\
&&D\varphi_3=d\varphi_3+d\psi-\frac{1}{\Delta}s^4_\theta s^4_\alpha s^2_{2\beta}
s_\theta^2 c_\alpha^2 d\psi=d\phi_1-\frac{4\prod \mu_i^2}{\Delta}\frac{d\psi}{\mu_1^2}
\eea
Using these expressions, we can simplify the deformed solution  in \nref{Mthdeform}

\section{Classical solutions for the probe string.}
\renewcommand{\theequation}{B.\arabic{equation}}
\setcounter{equation}{0}

Lets us analyze the  motion of a probe string in the geometry
(\ref{GammaDeform}). We
look
for a solution of the form
\bea
\psi=\psi_1(\tau)+\psi_2(\sigma),\quad \varphi_1(\sigma),\quad
\varphi_2(\sigma),\quad t=ER^{-2}\tau
\eea
while coordinates $\alpha$ and $\theta$ are constant.
With this ansatz we have to impose the constraints
\bea
g_{\psi\mu}(x^\mu)'=0,\qquad R^{-2} E^2=g_{\psi\psi}{\dot\psi}^2+
g_{\mu\nu}(x^\mu)' (x^\nu)'
\eea
Then we find
\bea\label{ProbeEqn1}
G^{-1}R^{-4}(E^2-J^2)&=&s_\alpha^2(4-(12-8c_{2\theta}^2)s_\alpha^2+9s_{2\theta}^2
s^4_\alpha)\left\{
(1+9{\hat\gamma}^2 s^2_{\theta} c^2_\theta s_\alpha^4 c^2_\alpha)
A^2\right.\nonumber\\
&&\left.+\frac{s_{2\theta}^2[1+{\hat\gamma}^2
s_\alpha^2(c_\alpha^2+s_\alpha^2 s_{\theta}^2
c_{\theta}^2)]^2}{4c_{2\theta}^2c^2_\alpha(1+9{\hat\gamma}^2s_{\theta}^2
c_\theta^2
s_\alpha^4c_\alpha^2)}B^2\right\}
\eea
We  have introduced coefficients $A$ and $B$ which are related to various
quantities
in the following way
\bea
&&\psi'=-\frac{4c_\alpha^2-8c_{2\theta}^2
s_\alpha^2+9s_{2\theta}^2s_\alpha^4}{
6c_{2\theta}c_\alpha^2(1+9{\hat\gamma}^2s_{\theta}^2 c_{\theta}^2
s_\alpha^4c_\alpha^2)}B,
\nonumber\\
&&
\varphi_1'=2c_{2\theta}s_\alpha^2
A-c_{2\theta}s_\alpha^2{\hat\gamma}{\dot\psi}-
\frac{c_{2\theta}^2s_\alpha^2+2-3s_\alpha^2}{3c_{2\theta}c^2_\alpha}B\nonumber\\
&&
\varphi_2'=(c^2_{\alpha}-s_\alpha^2c^2_\theta)(2A-{\hat\gamma}{\dot\psi})+B
\left[1+\frac{c_{2\theta}^2s_\alpha^2+2-3s_\alpha^2}{6c_{2\theta}c^2_\alpha}\right
]\nonumber
\\
&&R^{-2}J={\dot\psi}-\frac{\hat\gamma}{2} s^2_\alpha
AG(4-(12-8c_{2\theta}^2)s_\alpha^2+
9s_{2\theta}^2 s^4_\alpha)
\eea
The advantage of writing (\ref{ProbeEqn1}) in terms of $A$ and $B$ is that
the expression
(\ref{ProbeEqn1}) is positive definite and for generic values of
$\alpha,\theta$ the BPS
condition leads to $A=B=0$ which translates into
\bea\label{PrimSolut}
\psi'=0,\qquad
\varphi_1'=-c_{2\theta}s_\alpha^2{\hat\gamma}{\dot\psi},\qquad
\varphi_2'=-{\hat\gamma}(c^2_{\alpha}-s_\alpha^2c^2_\theta){\dot\psi}
\eea
One can also check that equations of motion for $\alpha$ and $\theta$ are
satisfied. Let us
look at the angular momenta corresponding to the solution
(\ref{PrimSolut})
\bea
J_{\varphi_1}=-(c_\alpha^2-s^2_\alpha
c_{2\theta})J=R^2\frac{\varphi_2'}{\hat\gamma},\quad
J_{\varphi_2}=s_\alpha^2
c_{2\theta}J=-R^2\frac{\varphi_1'}{\hat\gamma},\qquad J_\psi=R^2{\dot\psi}
 \la{hundred}\eea
For closed string we need that
\bea
\varphi_1'=n_1,\quad \varphi_2'=n_2 ~,~~~~~~~n_{1,2} \in Z
\eea
since at generic values of $\alpha , ~\theta$ the two circles have a non-vanishing size.
In terms of these integers we find
\bea
J_{\varphi_1}=\frac{n_2}{\gamma},\quad J_{\varphi_2}=-\frac{n_1}{\gamma}.
\nonumber
\eea
These can only be integers if $\gamma $ is a rational number. When $\gamma = m/n$,
then we see that the momenta $J_{\varphi_i}$ are multiples of $n$.
We will also need the expressions for three angular momenta
$(J_1,J_2,J_3)$ which are
related to the quantities introduced above
\bea
J=J_1+J_2+J_3,\qquad J_{\varphi_1}=J_2-J_1,\qquad J_{\varphi_2}=J_2-J_3
\eea
So finally we find the expressions
\bea
J_3=J_2+\frac{n_1}{\gamma},\qquad J_2=J_1+\frac{n_2}{\gamma}
\eea
Note also that the relation \nref{hundred}
between the values of $\alpha, \theta$ and the $J_i$ is
such that
\be
 (|J_1|,|J_2|,|J_3|) = \lambda  ( \mu_1^2 , \mu_2^2 , \mu_3^2)
 \ee
for some $\lambda$.

\end{document}